\documentclass[11pt,preprint]{aastex}

\shorttitle{CONSTRAINTS ON A LOCAL GROUP X-RAY HALO}
\shortauthors{RASMUSSEN \& PEDERSEN}

\begin{document}

\title{CONSTRAINTS ON A LOCAL GROUP X-RAY HALO}

\author{Jesper Rasmussen\altaffilmark{1}}
\affil{Astronomical Observatory, University of Copenhagen, Juliane 
Maries Vej 30, DK-2100 Copenhagen \O, Denmark.}
\email{jr@astro.ku.dk}
\and
\author{Kristian Pedersen\altaffilmark{2}}
\affil{Danish Space Research Institute, Juliane 
Maries Vej 30, DK-2100 Copenhagen \O, Denmark.}
\altaffiltext{1}{also at: Danish Space Research Institute, Juliane 
Maries Vej 30, DK-2100 Copenhagen \O, Denmark.}
\altaffiltext{2}{present address: Astronomical Observatory, University of 
Copenhagen, Juliane Maries Vej 30, DK-2100 Copenhagen \O, Denmark; 
kp@astro.ku.dk.}

\begin{abstract}
A simple model for a hot Local Group halo is constructed, using the
standard $\beta$-model for the halo density and by choosing model parameters
based on all available observations of X-ray emission in other groups of 
galaxies and on optical data on Local Group morphology. From the 
predicted X-ray intensities, total Local Group mass, and central cooling time
of the halo, we derive very conservative upper limits on the central halo 
density $N_0$ and global temperature $T$ of $N_0=5\times10^{-4}$ cm$^{-3}$ 
and $kT=0.5$ keV, irrespective of realistic values of the density profile 
parameters $r_c$ and $\beta$. A typical poor group value of $\beta=0.5$
requires $kT<0.15$ keV and $N_0<10^{-4}$ cm$^{-3}$, from which 
it is concluded that the Local Group is very unlikely to possess a 
significant X-ray halo. The prospects for further constraining of halo 
parameters from UV absorption line observations are considered. We 
explicitly calculate the
ability of the halo to distort the cosmic microwave background (CMB) in terms
of the resulting CMB temperature variations and multipole anisotropies. 
\end{abstract}


\section{INTRODUCTION}\label{sec,intro}
It is well established \cite{mulc1996b,mahd2000,hels2000} that some groups of 
galaxies contain an X-ray emitting intragroup medium. Thus, one might also
suspect the Local Group to possess hot, intergalactic gas. The significance
of such a Local Group X-ray halo, however, has been a matter of debate, and 
it has even been suggested \cite{suto1996} and subsequently rejected
\cite{band1996,pild1996} that this halo might
influence the observed microwave background anisotropies through the
Sunyaev-Zel'dovich effect.

Previous models of intergalactic gas distributed in a Local Group halo 
and/or its possible diffuse X-ray emission include those of Kahn \& Woltjer 
(1959) and Oort (1970); Hunt \& Sciama (1972); Suto et al.\ (1996), and 
related works by Banday \& Gorski (1996) and Pildis \& McGaugh (1996); 
Bland-Hawthorn (1999) and Maloney \& Bland-Hawthorn (1999);  Sidher, 
Sumner, \& Quenby (1999); and a recent observational study by 
Osone et al.\ (2000).

Here we present a new model based on the standard $\beta$-model for the 
density of halo gas, but with more realistic parameters for the gas and for
the Local Group itself than previously presented in the literature. The
choice of input parameters is based on recent optical observations of 
the Local Group as well as on all available X-ray observations of other
groups and clusters of galaxies. We examine the constraints imposed on a 
Local Group X-ray halo by
observations of the soft X-ray background, estimates of Local Group mass,
and by the fact that the existence of a present-day hot halo requires it to
be stable towards cooling over a significant fraction of the Hubble time 
$t_H$.

The paper is organized as follows. In \S~\ref{sec,model} we present the model 
halo, with
particular emphasis on its deviations from earlier models. In \S~\ref{sec,obs}
we discuss
the observational (X-ray and optical) constraints on the model. Model results
for X-ray intensities, gravitating mass, and cooling time scale are presented
in \S~\ref{sec,results} and compared with observations in 
\S~\ref{sec,discus}, from which 
revised model parameters are extracted. Features of the new model, including 
its ability to absorb distant UV radiation and its effect on the Cosmic 
Microwave Background (CMB hereafter), are given in \S~\ref{sec,newmodel}, and 
conclusions are presented in \S~\ref{sec,conclude}.

\section{MODELLING THE LOCAL GROUP HALO}\label{sec,model}

\subsection{Density Profile}
The Local Group Halo (LGH hereafter) is here assumed to be spherically
symmetric, isothermal, and described by the standard
$\beta$-model of Cavaliere \& Fusco-Femiano (1976),
\begin{equation}
N(r)=N_{0}\left[1+\left(\frac{r}{r_{c}}\right)^{2}\right]^{-3\beta/2},
\label{eq,beta}
\end{equation}
where $N_0$ is the central electron number density, $r$ is radial
distance, $r_c$ is the core radius,
and $\beta$ is a dimensionless parameter. This gas profile provides a good
fit to the X-ray emission of galaxy groups and is therefore often assumed
for the spatial distribution of gas in groups with an X-ray-detectable
intragroup medium \cite{mulc1996b,hels2000}.

While presenting our model of the LGH, we will briefly recapitulate important
characteristics of previous models and the associated results. In this 
context, the first three papers mentioned in \S~\ref{sec,intro} should be 
treated separately. It suffices to mention that
Kahn \& Woltjer (1959) suggested a homogeneous temperature and density of 
Local Group intergalactic gas of $\sim 0.04$ keV and $\sim 10^{-4}$ cm$^{-3}$,
respectively. In their model, and in that of Oort (1970), this gas
is required to dynamically stabilize the Local Group, but the argument is 
based on the then known masses of M31 and the Galaxy of $\sim10$\% of the 
currently accepted values. Hunt \& Sciama (1972) estimate the LGH temperature 
and density to be $\sim 0.02-0.25$ keV and $\sim 2-5\times10^{-4}$ cm$^{-3}$
from the now obsolete assumption that the observed soft X-ray background at
$E=0.28$ keV could largely arise from an enhancement of X-ray emission from 
the LGH gas due to the 
gravitational influence of the Galaxy moving through this intergalactic gas.

Newer models all assume the X-ray halo to be spherically symmetric and 
isothermal, and all assume $\beta=2/3$ in equation~(\ref{eq,beta}), with the 
exception of Sidher et al.\ (1999) who assume 
an $l/r^2$-behavior, where $l$ is the distance along the line of sight.
We note, however, that while $\beta=2/3$ appears to be representative of 
clusters of galaxies (e.g., Jones \& Forman 1984), poor groups tend to have 
shallower gas profiles, with $\beta \sim 0.5$ 
\cite{mulc1996b,dahl2000,hels2000}, which 
will thus act as our initial model value. Regarding the other density 
structure parameter, $r_c$, Suto et al.\ (1996) use $r_c=150$ kpc as does 
Bland-Hawthorn (1999),
while Maloney \& Bland-Hawthorn (1999) consider the range 
$r_c=[0, 350 \mbox{ kpc}]$.
Banday \& Gorski (1996) and Osone et al.\ (2000) only restrict themselves to
core radii smaller than the distance from the observer to the LGH center. 
For a relatively poor group like the Local Group, a value of
$r_c=150$ kpc (again typical of clusters) seems rather large when viewed in 
light of observational results on other groups 
(Mulchaey et al.\ 1996b; Pedersen, Yoshii, \& 
Sommer-Larsen 1997; Dahlem \& Thiering 2000).
In contrast, Pildis \& McGaugh (1996) choose $r_c=33$ kpc, which appears more
reasonable as a starting point. 
Finally, for the central density $N_0$, values both
above and below $10^{-3}$ cm$^{-3}$ have been used in LGH models. Typical
gas densities in clusters are of this magnitude, which may therefore apply 
also to the central parts of groups, as confirmed by the results of 
e.g., Davis et al.\ (1995), and Dahlem \& Thiering (2000). Hence, this value 
of $N_0$ is initially implemented in our model.

\subsection{Temperature}
From X-ray observations of other poor groups of galaxies, intragroup plasma
temperatures for groups containing at least one elliptical galaxy are 
typically inferred to be $\sim 1$ keV 
\cite{mulc1996b,tawa1998,davi1999,hels2000}, which is 
also the value used in the models of Suto et al.\ (1996) and
Pildis \& McGaugh (1996). A more obvious choice for a LG model would be
the virial temperature $T_{vir}$ of any hot Local Group medium. Strictly 
defined, $T_{vir}=\mu m_p \sigma^2/(3k)$,
where $\sigma$ is the 3-D velocity dispersion of ions in the halo gas,
$m_p$ is the proton mass, and $\mu$ is the mean molecular weight in amu. 
Using the radial velocity dispersion of Local Group galaxies of
$\sigma_r=61\pm8$ km s$^{-1}$ \cite{vand2000}, we obtain 
$T_{vir}\simeq 0.03$ keV. On the other hand, according to Maloney \& 
Bland-Hawthorn (1999) and to the result that would be obtained for 
$T_{vir}$ by Mulchaey et al.\ (1996b) using the above value of $\sigma_r$, 
$T_{vir} \approx 0.2$ keV. Thus, the model constraint of 
$N_0r_c \leq 1.5\times10^{21}$ cm$^{-2}$ of Maloney \& Bland-Hawthorn 
applies at $T=0.2$ keV. 
However, since any estimate of the virial temperature is based on a 
number of assumptions, in particular that the the velocity ellipsoid of the
galaxies is isotropic, and that the typical ion velocity is 
comparable to that of the galaxies, it is instructive to consider a range of 
realistic temperatures for a hot Local Group halo (the former assumption 
has for example not yet been justified for Local Group galaxies). Further,
due to the pronounced substructure exhibited by the Local Group luminous 
mass, $\sigma_r$ may not be a reliable estimator of the Local Group 
gravitational
potential, and hence, virial temperature. Therefore, we will consider all 
temperatures in the range $\sim0.03$ to $\sim1$ keV, i.e.\ on the order of 
the above virial temperature and 
up to a value representative of poor groups with an X-ray detectable 
intragroup medium. Although the Local Group does not contain any major
elliptical galaxies (the only elliptical being the E2 dwarf M32 with a 
$V$-magnitude of $M_V=-16.5$ as compared to, e.g., the Small Magellanic 
Cloud with $M_V=-17.1$; van den Bergh 2000) we will, however, as a 
starting point use the above mentioned typical group temperature of 
$\sim 1$ keV.

\subsection{Metallicity}
Regarding the metallicity $Z$ of halo gas, $Z=0.3Z_{\odot}$ is used in all 
previous models when modelling the LGH X-ray emission 
\cite{suto1996,oson2000}, although Maloney \& Bland-Hawthorn (1999) also 
consider
$Z=0.01Z_{\odot}$ and $Z=0.1Z_{\odot}$. The value  $Z=0.3Z_{\odot}$, 
however, is typical of clusters and rich groups of galaxies (e.g.,
Hwang et al.\ 1999) whereas poor low-temperature ($T\lesssim 1-1.5$ keV) 
groups tend 
to have somewhat lower metallicities, possibly because their gravitational 
potentials are too shallow to retain all the enriched ejecta of the 
constituent galaxies 
\cite{tawa1998,davi1999,hwan1999}. This view is supported by the 
gas-to-total-mass ratios 
of such groups being lower than in rich groups and clusters 
\cite{david1995,renz1997}. As the LG is certainly a relatively poor 
group, $Z\lesssim 0.2 Z_{\odot}$ for the LG gas seems more reasonable. This 
upper limit on $Z$ will henceforth be adopted as our model value.

\subsection{Observer Position}
\begin{figure}
\epsscale{.6}
\plotone{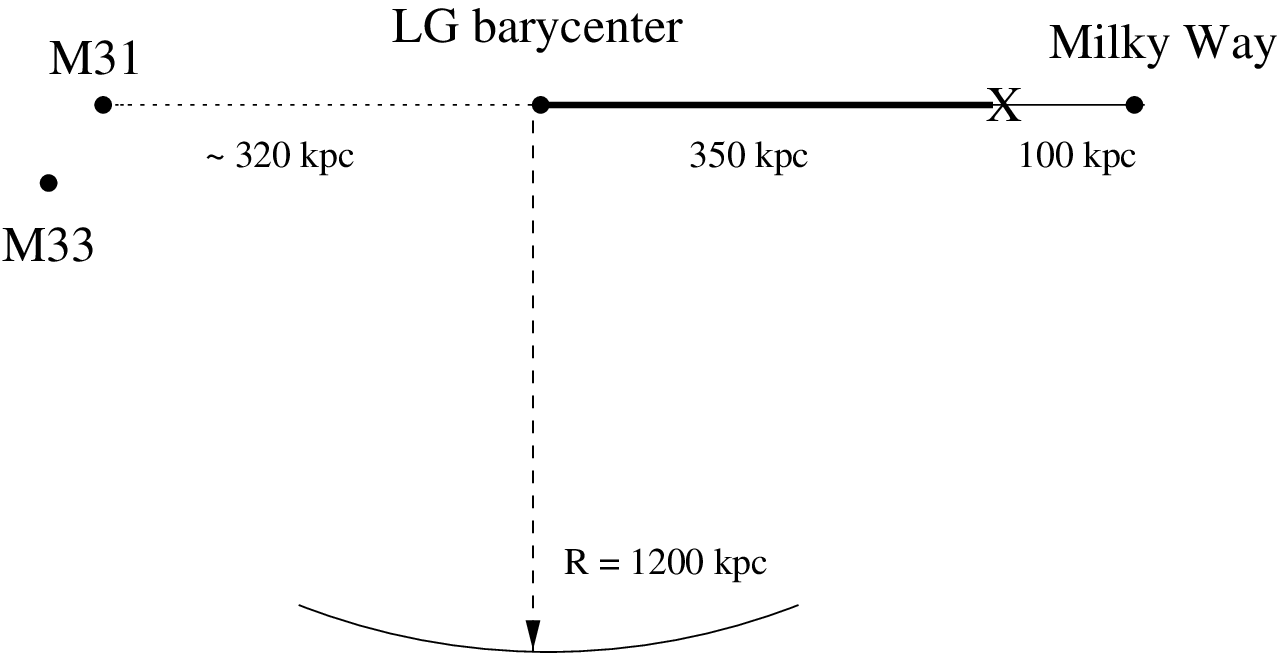}
\caption{Adopted Local Group geometry. M31, the Local Group Barycenter
(LGBC), and the Milky Way (MW) are assumed to be aligned. The 'X' marks the 
often adopted MW-LGBC distance of $x_0=350$ kpc (see text), while the thick 
line corresponds to the range in $x_0$ considered by 
Maloney \& Bland-Hawthorn (1999). Here we assume $x_0=450$ kpc, while 
noting that taking $x_0 \lesssim 350$ kpc implies that the MW subgroup should
be more massive than the M31 subgroup for a fixed MW-M31 distance of 
$\sim 770$ kpc. The dashed line illustrates the distance to the cut-off 
surface at $R=1200$ kpc, beyond which the halo gas density is assumed to be 
zero. \label{fig1}}
\end{figure}

A final vital parameter is the distance $x_0$ from the observer to the LGH 
center.
Suto et al.\ (1996) use $x_0=0$ when modelling the halo X-ray emission, and 
claim on the basis hereof that the LGH may contribute a non-negligible part 
to the low-energy ($E \lesssim 1.5$ keV) X-ray Background, but without 
taking into account 
Galactic foreground absorption. They use $x_0=350$ kpc, when estimating the 
upper limit on the LGH-induced Sunyaev-Zel'dovich effect in {\em COBE} data. 
This 
value is also adopted in the model of Sidher et al.\ (1999), while Maloney \& 
Bland-Hawthorn (1999) again consider the entire range 
$x_0=[0, 350 \mbox{ kpc}]$. 
We may, however, obtain a realistic estimate of $x_0$ through the following
considerations.
Morphologically the Local Group exhibits a bimodal spatial structure in the
optical, consisting of the M31 and Milky Way subgroups. As these two
subsystems account for the large majority of Local Group mass 
(Zaritsky 1994, 1999; Mateo 1998; Courteau \& van den Bergh 1999),
the Local Group Barycenter (LGBC hereafter) should be situated close to the 
line connecting M31 and the Milky Way, i.e.\ in galactic direction 
($\ell ,b)=(121^{\circ}\!\!.7,-21^{\circ}\!\!.3)$, with a possible, however
small, offset towards M33 at ($133^{\circ}\!\!.6,-31^{\circ}\!\!.5)$.
Here we will identify the center of the Local Group halo with the LGBC, taken
to be in the direction of M31 (see Figure~\ref{fig1} for a schematic view
of the assumed Local Group geometry).
The distance to M31 appears to be rather well-determined to 0.76--0.78 Mpc
\cite{kara1996,stan1998,vand2000}. 
Thus, choosing $x_0 \leq 350$ kpc implies that the Milky Way 
subgroup should be more massive than that of M31. However, Zaritsky (1999) 
applies the simple timing 
argument of Kahn \& Woltjer (1959) to infer a mass ratio of 1.5 for the
M31--Milky Way pair. Van den Bergh (2000) uses this ratio together with a 
distance $d$ to M31 of 0.76 Mpc to suggest a value of 
$0.6d\simeq 450$ kpc for the distance to the LGBC. This value is 
adopted here for $x_0$, which is 100 kpc larger than typically assumed in 
previous LGH models. The inclusion of M33 in this estimate only serves to
widen the gap between this and previous models.

The model halo is cut off at the zero-velocity surface $R$ separating the 
gravitational contraction of the LG from the Hubble expansion. This is
estimated by Courteau \& van den Bergh (1999) to $R=1.18\pm 0.15$ Mpc, 
based on which we here take $R=1200$ kpc, which is
also the radius to which reliable dynamical mass estimates of the LG are 
obtainable \cite{zari1994}.

\placefigure{fig1}

\subsection{X-ray Emission and Absorption}
For the model plasma, intensity calculations are done by means of the {\sc 
meka} emissivity code 
(Mewe, Gronenschild, \& van den Oord 1985; Mewe, Lemen, \&  van den Oord 1986)
in a revised version last updated 
1991 by J.S.\ Kaastra. To compare model results with X-ray background (XRB
hereafter) 
measurements, the {\em ROSAT} or {\em ASCA} energy resolution should be used 
in the code.
The {\em ROSAT} PSPC energy resolution approximates
$\delta E/E =0.43(E/0.93)^{-0.5}$ (FWHM) over the entire detector area, 
i.e.\ $\sim0.4$ keV at $E=1$ keV,
while the corresponding {\em ASCA} GIS resolution is
$\delta E/E =0.079 \sqrt{5.9/E}$, i.e.\ $\sim0.2$ keV.
In the calculations the better {\em ASCA} resolution is used, but note that 
{\em ASCA} 1 keV model intensities are $\sim 20$\% higher than those of 
{\em ROSAT} at 
$T=1$ keV (for a plasma with $Z=0.2Z_{\odot}$), 
and in the range $T\simeq 0.2-0.5$ keV there is even a factor of 2 
difference, with {\em ROSAT} model intensities being the largest.

Regarding the effects of absorption, the neutral hydrogen column density 
$N_H$ in the direction of M31 is 
$\sim7\times 10^{20}$ cm$^{-2}$ \cite{star1992}.
However, conservatively selecting a field of interest of 
$10^{\circ} \times 10^{\circ}$ centered on M31 (this will be justified 
in \S~\ref{sec,obs}), we expect maximal
absorption within this field to occur closest to the Galactic plane, where
$N_H \simeq 10^{21}$ cm$^{-2}$. Using this value along with the absorption
cross section of Morrison \& McCammon (1983), we find a maximal absorption
in the selected field of $\sim 20$\% at $E=1$ keV. This is comparable
to computational (as shown above) and observational (see \S~\ref{sec,obs}) 
uncertainties on the 1 keV XRB intensity, even using this 
conservative absorption estimate. Since we will further
apply a conservative upper limit on the XRB normalization in 
\S~\ref{sec,obs}, we will for these reasons neglect absorption in the 
following.

\subsection{Model Summary}
To provide a starting point, a set of ``reference values'' is chosen as 
initial input parameters for the halo gas. Based on the above discussion, we 
take the initial plasma parameters to be 
$N'_0=10^{-3}$ cm$^{-3}$, $r'_c=20$ kpc, $\beta'=0.5$, 
$Z'=0.2 Z_{\odot}$, and $T'=1$ keV, with $x_0=450$ kpc and $R=1200$ kpc. 
We stress that these parameters are selected based on all available X-ray 
observations of other groups of galaxies as well as corresponding cluster 
observations and optical observations of Local Group galaxies.

\section{OBSERVATIONAL CONSTRAINTS}\label{sec,obs}

Recent observations of the XRB have been performed both 
in the form of pointed observations, mainly in directions with low 
Galactic absorbing column density, and in the form of the {\em ROSAT} All-Sky 
Survey.
After correction for Galactic absorption, the spectral shape of the 
extragalactic XRB in the $1-10$ keV energy range is well fitted 
by a power-law, $I(E)=AE^{-\alpha}$, with a spectral
normalization $A$ at $E=1$ keV of 
$\sim 10-11$ keV cm$^{-2}$ s$^{-1}$ sr$^{-1}$ keV$^{-1}$ inferred from pointed 
observations
(e.g., Garmire et al.\ 1992; Chen, Fabian, \& Gendreau 1997; 
Miyaji et al.\ 1998; Vecchi et al.\ 1999).

For our LGH model we expect the largest intensities in the direction of the 
adopted LGH center, towards M31 at $(121^{\circ}\!\!.7,-21^{\circ}\!\!.3)$.
M31 itself has an angular extent of $\sim 3^{\circ}$, and most previous X-ray 
studies in directions towards M31 have aimed at detecting and 
classifying discrete sources or regions within
the galaxy and have therefore been limited to comparatively small fields. 
We note, however, that for an isothermal bremsstrahlung model LGH with our
model values, the halo intensity has fallen to 25\% of peak value already at 
$5^{\circ}$ from the direction to the halo center.
Since the disk (West, Barber, \& Folgheraiter 1997) and bulge (Primini, 
Forman, \& Jones 1993) of M31 are themselves sources of diffuse soft
X-ray emission (although the fraction of bulge emission attributable to hot 
gas is still debated, see, e.g., Irwin \& Bregman 1999;  
Borozdin \& Priedhorsky 2000; Primini et al.\ 2000; Shirey et al.\ 2000), 
emission from a LGH centered directly in front 
of M31 would thus be hard to disentangle from the combined emission of M31 
and the extragalactic XRB. Since any search for 
non-galactic diffuse X-ray emission associated with 
the LGH at positions close to M31 will necessarily be 
hampered by emission from M31, and since the LGH center may be 
slightly offset from the direction of M31 with some unknown amount, this
suggests that searching for any direct evidence of LGH emission should be 
done in a field of several tens of square degrees.

The {\em ROSAT} all-sky maps of Snowden et al.\ (1997) in the R6 
($0.9-1.3$ keV) 
band do not show clear signs of degree--scale excess extragalactic diffuse
emission in a field of $10^{\circ} \times 10^{\circ}$ centered on M31, 
however. These all-sky maps have been removed of point sources to a uniform 
flux level for which the original survey source catalog (RASS-I) was 
complete over 90\% of the sky. To obtain a qualitative picture of the amount 
of any excess 1 keV emission towards M31, we resampled the original 
``point source--removed'' $102^{\circ}$ by $102^{\circ}$ image centered at 
($\ell ,b)=(90^{\circ},0^{\circ}$)
into pixels of $0.5^{\circ}$ at each side, smoothed it with a 
box of $3\times 3$ pixels, and added the 
photon count rates at each longitude in
the resulting image within a $10^{\circ}$ latitude bin centered at 
$b=-21^{\circ}$.
Again, the result shows no evidence of excess diffuse emission within a
$10^{\circ} \times 10^{\circ}$ field centered on M31 as compared to other
fields at similar latitude. 
Centering the strip at $b=-27^{\circ}$ instead, to include the latitude
of M33 and the region covered by Osone et al.\ (2000) (see below), does
not change this conclusion.

A recent, much more detailed investigation by Osone et al.\ (2000), 
to our knowledge the first one to 
be carried out with the distinct purpose of constraining LGH parameters from
direct observations of associated X-ray emission, 
concentrated on 4 neighboring {\em ASCA} GIS pointings centered approximately
$6^{\circ}$ south of M31 (i.e.\ a few degrees off the line connecting
M31 and M33 as seen from the Sun). For reference purposes spectral data for 
a region close to the 
North Galactic Pole was used. From fits to the observed spectra the authors 
find no evidence of any soft (0.6-2 keV) excess flux in these directions 
relative to the general XRB as 
represented in their reference direction. They place an upper limit on this
excess flux of 
$1.27\times10^{-8}$ erg cm$^{-2}$ s$^{-1}$ sr$^{-1}$ between 0.6 and 2 keV,
from which they conclude that
$[N_0/10^{-3}$ cm$^{-3}]^2[r_c/100$ kpc$] \leq 5.80\times10^{-3}$ by
modelling the LGH plasma with a $\beta$-model of $\beta=2/3$ and its emission
with a Raymond-Smith plasma code of $Z=0.3Z_{\odot}$ and $T=0.3-1.2$ keV.
An important point about the result of Osone et al.\ is that 
although their pointing direction
ensures reasonably low Galactic absorption and minimization of M31 
contamination, the sky coverage is limited to $\sim 5$ deg$^2$. Thus, they 
may very well have missed the true LGH center by a few degrees, implying
a possible underestimation of the upper limit on the quantity $N_0^2r_c$.
From the result of Osone et al., and from our own quick-look analysis
mentioned above, we infer typical unabsorbed intensities
in the LGBC direction not to be in significant excess of 12 keV cm$^{-2}$ 
s$^{-1}$ sr$^{-1}$ keV$^{-1}$. Neglecting absorption, this value will thus be 
invoked initially as a maximally allowed model normalization to constrain 
possible LGH 
parameter combinations. Note that we hence allow the halo to be solely 
responsible for the observed intensities in the relevant direction. This 
should yield conservative limits on halo parameters.

In order to further constrain parameters, the central cooling time
scale $t_c$ of the LGH plasma will also be considered. For the LGH to
exist and be reasonably stable towards cooling, we take the condition 
$t_c > t_H/2$ with $t_H =15$ Gyr. If this cooling stability 
criterion is not fulfilled, the LGH would have
cooled substantially during its lifetime, in which case there would
hardly be any present-day hot halo of the LG. This scenario is entirely 
plausible, but since we want to pinpoint the present {\em upper limits} on 
central
density and temperature of the halo, we should require the model halo not to
have cooled away, i.e.\ be moderately stable towards cooling. Alternatively, 
the cooling gas 
would need to be replenished as also noted by Bland-Hawthorn (1999). As there
are no galaxies within the central $\sim300$ kpc of the LGBC and since
Local Group dynamics \cite{peeb1990,zari1994} indicate that the 
M31--Galaxy pair is approaching each other for the first time, 
standard mechanisms like galactic 
winds or ram pressure stripping would seem incapable of providing
the required additional X-ray gas to a cooling central halo region.

Finally, we note that dynamical estimates of total LG mass have 
approached a recent consensus towards the value $\sim 3\times 10^{12} 
M_{\odot}$ \cite{byrd1994,zari1994,mate1998,cour1999}. This number 
will also be used in assessing the allowed halo parameter space.

\section{MODEL RESULTS}\label{sec,results}

\begin{figure}
\epsscale{1.0}
\plottwo{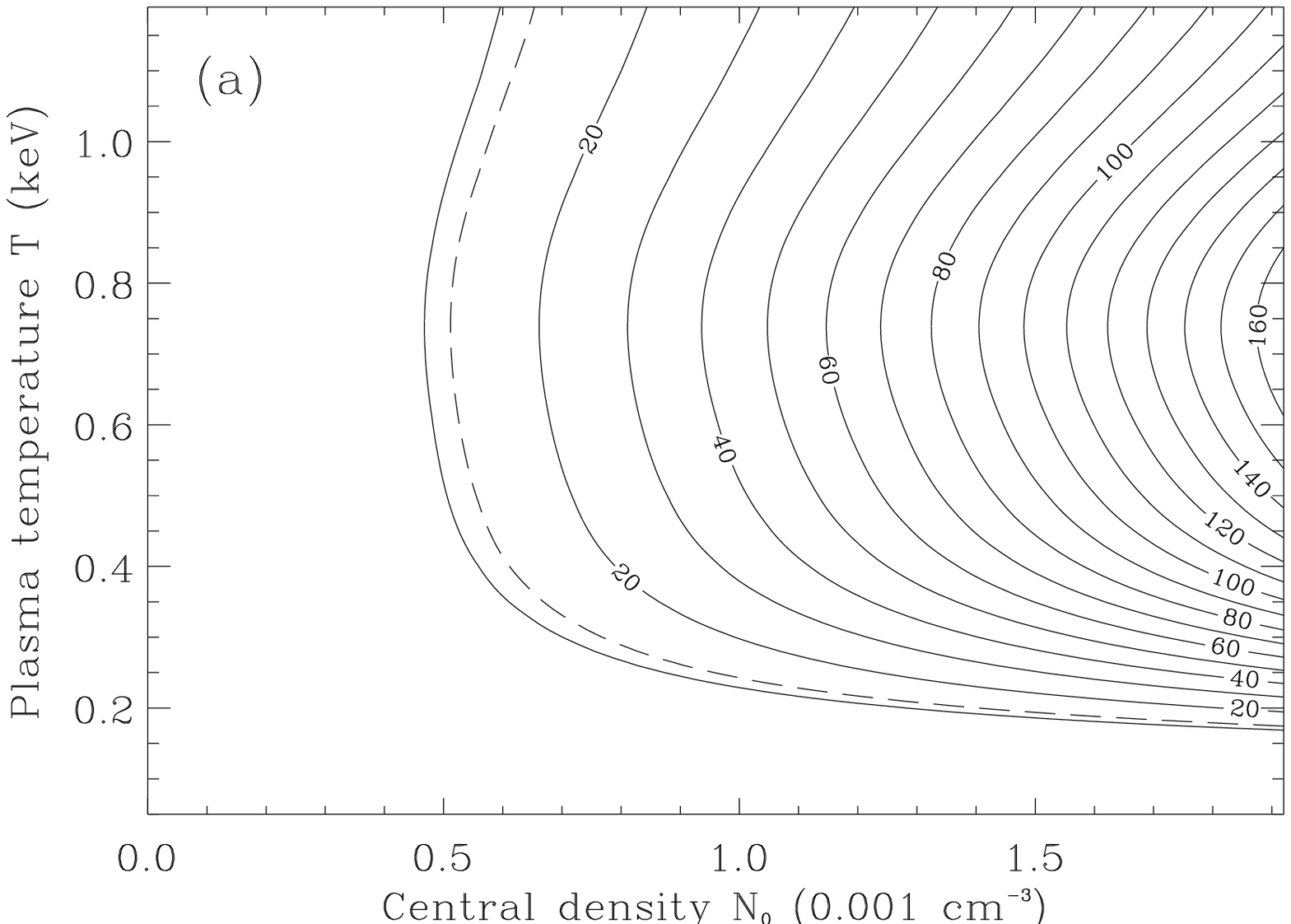}{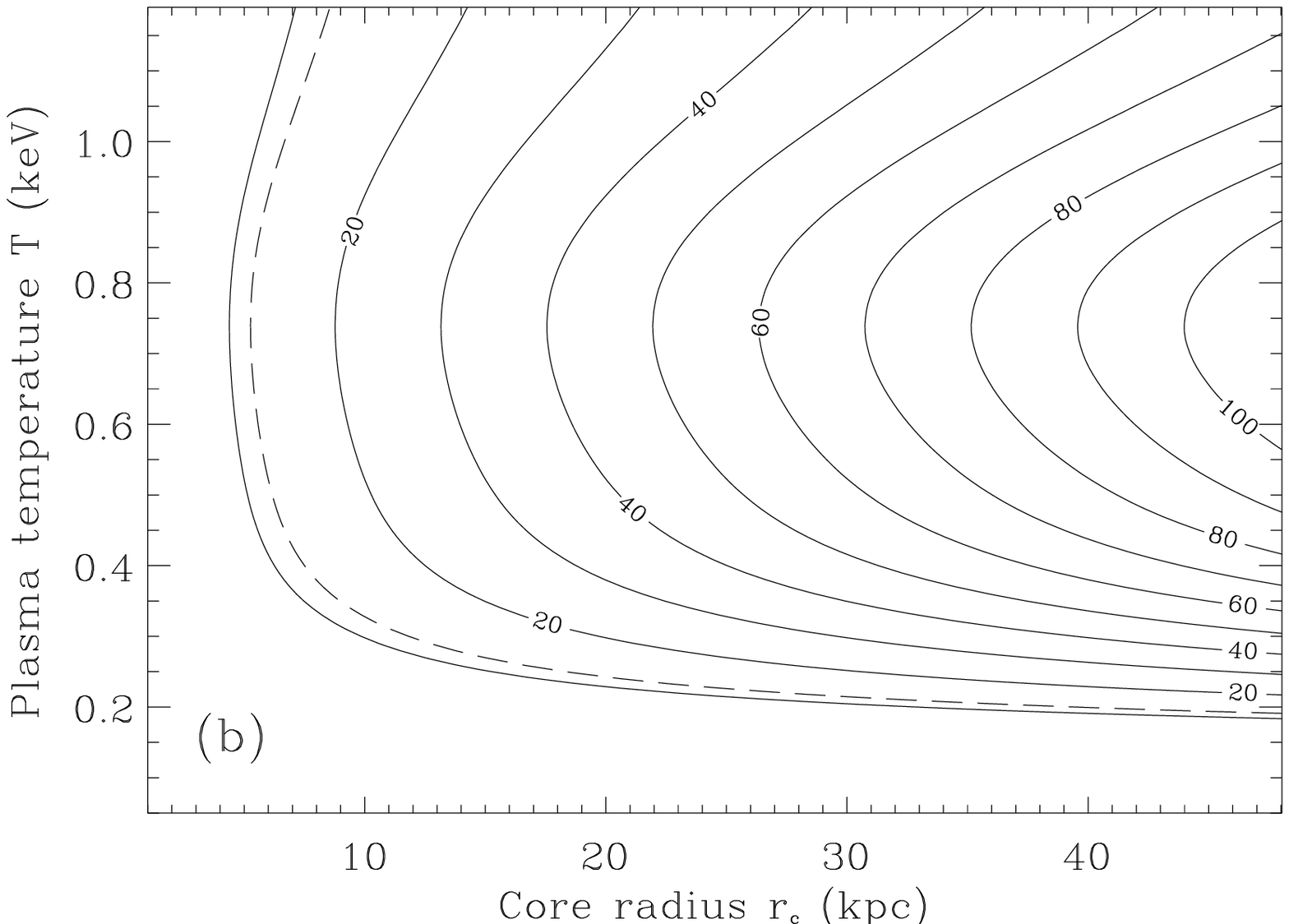}
\epsscale{2.22}
\plottwo{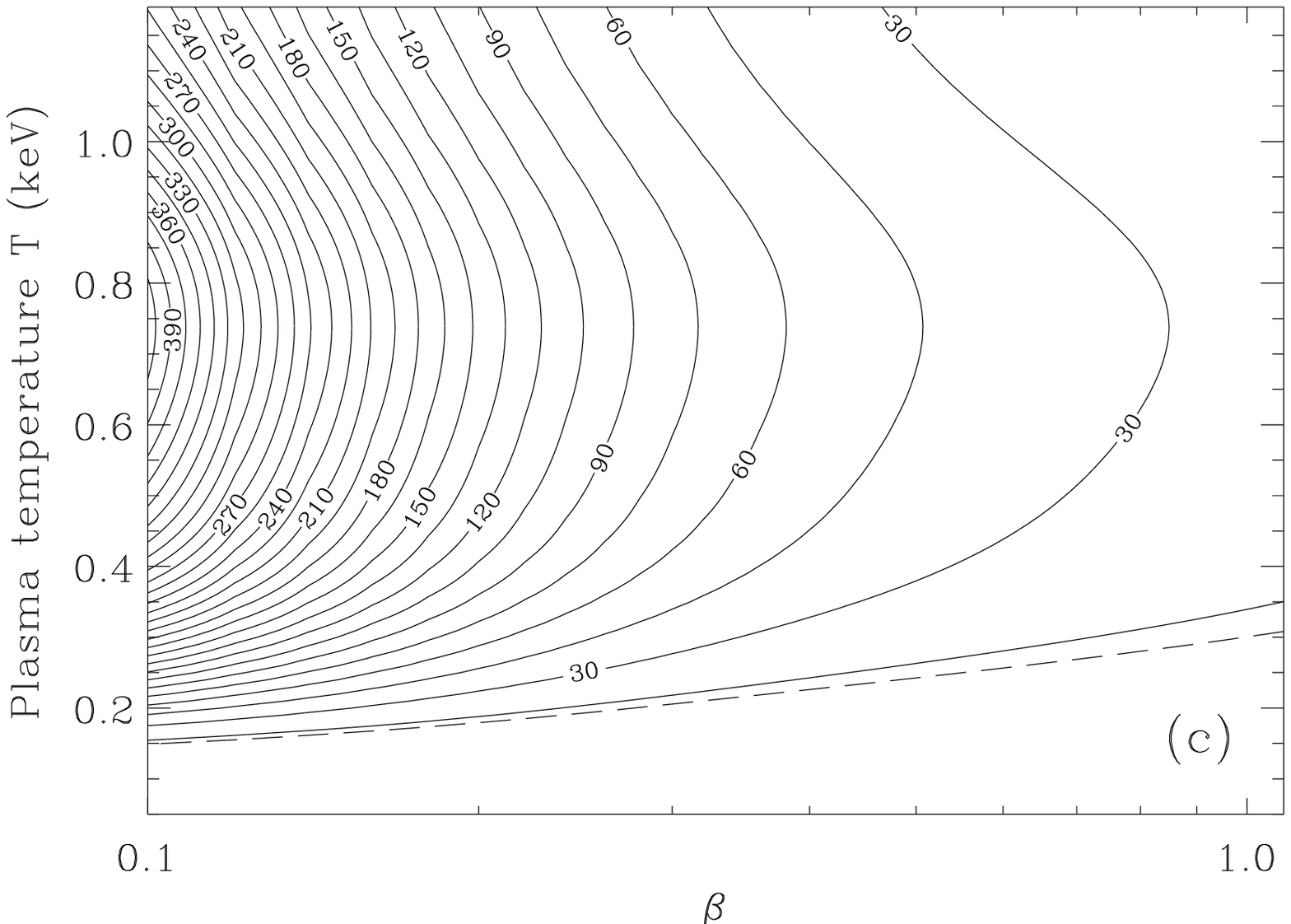}{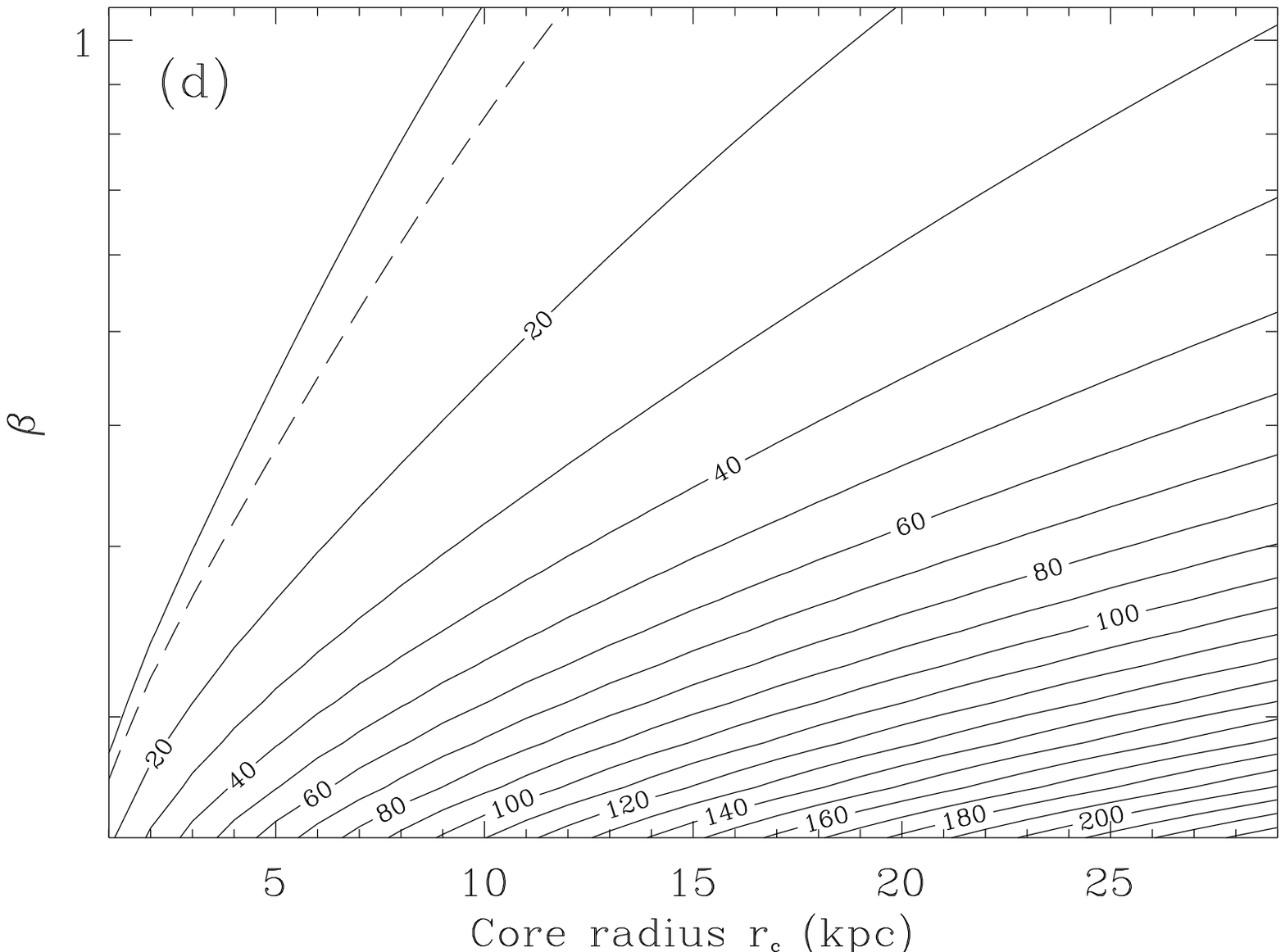}
\epsscale{4.94}
\plottwo{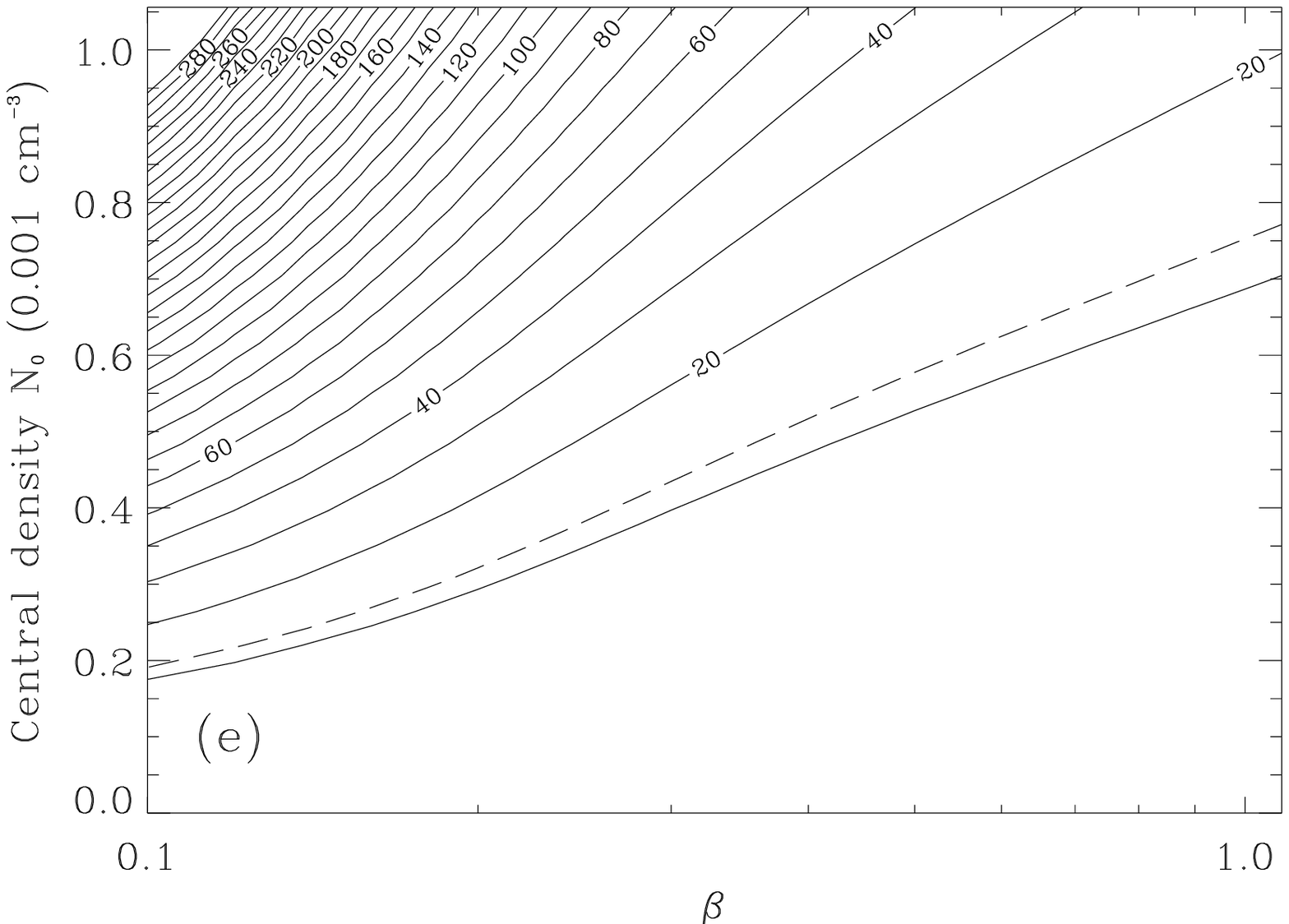}{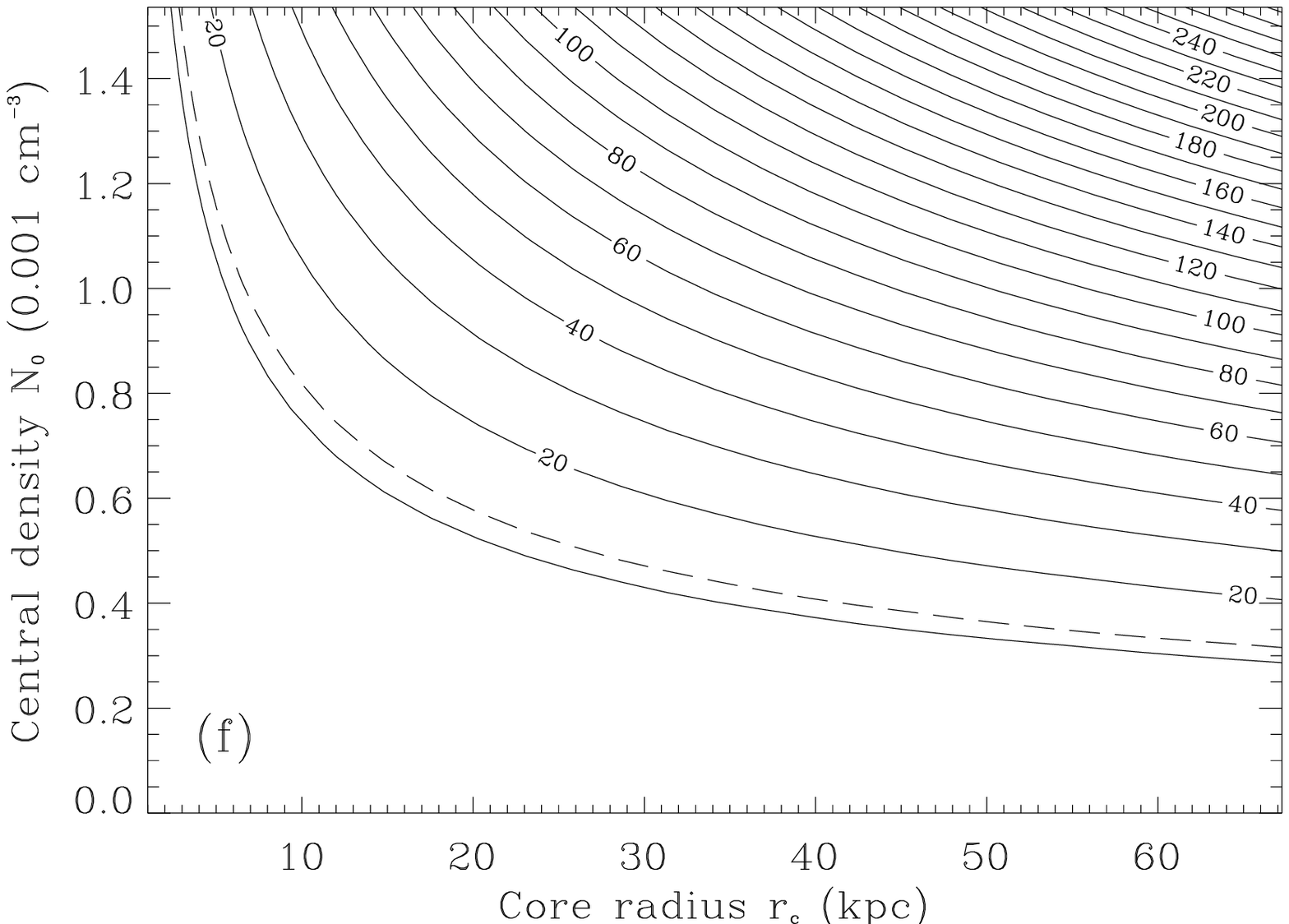}
\caption{Contours of predicted X-ray intensities of the LGH 
(units of keV cm$^{-2}$ s$^{-1}$ sr$^{-1}$ keV$^{-1}$) at $E=1$ keV as
function of $N_0$ and $T$ (a), $r_c$ and $T$ (b), $\beta$ and $T$ 
(c), $r_c$ and $\beta$ (d), $\beta$ and $N_0$ (e), and $r_c$ and $N_0$ (f). 
Viewing direction is
towards the LGH center at galactic coordinates ($\ell ,b)=
(121^{\circ}\!\!.7,-21^{\circ}\!\!.3)$. Dashed line shows the
inferred maximum value allowed by observations, see \S~\ref{sec,obs}. 
Fixed parameters are maintained at the values given 
in \S~\ref{sec,model}.\label{fig2}}
\end{figure}

Figure~\ref{fig2} presents results for X-ray intensities as calculated by
means of
the {\sc meka} code, with two parameters fixed at their reference values 
(given in \S~\ref{sec,model}) and
two parameters allowed to vary. Parameters inferred from 
optical observations ($x_0$ and $R$) remain fixed. Following Osone et al.\ 
(2000), temperatures up to 1.2 keV are considered. We restrict ourselves
to considering a photon energy of $E=1$ keV, as most XRB normalizations are 
given at this energy.
                                                                    
\placefigure{fig2}
\notetoeditor{Please join Figures ``\ref{fig2}a''--''\ref{fig2}f'' 
into ONE 
two-column three-row figure, denoted Figure~\ref{fig2} (Figure~\ref{fig2}a 
and \ref{fig2}b to the left and right, respectively, in the upper row, 
Figure~\ref{fig2}c and \ref{fig2}d to the left and right, respectively, 
in the middle row, and Figure~\ref{fig2}e and \ref{fig2}f to the 
left and right, respectively, in the lower row).}

The cooling time scale $t_c$ can be calculated given some cooling function
$\Lambda(T)$. Following Mushotzky (1993), we here invoke the relativistically
corrected Raymond-Smith cooling curve of Gehrels \& Williams (1993).
Noting that this $\Lambda(T)$ has a minimum of $\sim4\times 10^{-23}$
ergs cm$^3$ s$^{-1}$ in the soft X-ray regime ($10^6\lesssim T \lesssim 
10^7$ K), corresponding to a maximum in cooling time, we find
\begin{equation}\label{eq,tcool}
t_c \approx \frac{5kT}{N_e\Lambda(T)} < 
5.8 (T/10^7 \mbox{K})/(N_e/10^{-3} 
\mbox {cm$^{-3}$}) \mbox{ Gyr}
\end{equation}
as an upper limit on the actual cooling time. In Figure~\ref{fig3}, 
$t_c$ is plotted against $N_0$ and $T$ over the range within which 
equation~(\ref{eq,tcool}) is valid.

\placefigure{fig3}

Assuming our model LGH to be in hydrostatic equilibrium, we can also 
calculate its total gravitating mass 
$M_{grav}(r)$ residing within some radius $r$. If the LGH gas is supported 
against gravity solely through a 
thermal pressure given by the perfect gas law, one finds
\begin{equation}\label{eq,mass}
M_{grav}(r)=-\frac{kT_{gas}(r)r}{G\mu m_p} \left(\frac{d\, \mbox{ln}\, 
\rho_{gas}}{d\, \mbox{ln}\, r}+
\frac{d\, \mbox{ln}\, T_{gas}}{d\, \mbox{ln}\, r} \right) 
\end{equation}
for a gas distribution with mass density 
$\rho_{gas}$ (eq.~[\ref{eq,beta}]), mean molecular weight $\mu$, 
and with $m_p$ being the proton mass (Fabricant, Lecar, \& Gorenstein 1980). 
In Figure~\ref{fig4}, 
$M_{grav}$ within the adopted cut-off radius of $R=1200$ kpc is plotted
against $\beta$ and $T$, assuming $\mu=0.6$ (primordial 
abundances). Notice that only $\beta$ and $T$ remain free parameters
because $M_{grav}$ is independent of $N_0$ for a $\beta$-model gas 
distribution, and the results obtained from Figure~\ref{fig4} are 
insensitive to the choice of $r_c$,  as we will see
in the next section.

\placefigure{fig4}

\section{DISCUSSION}\label{sec,discus}

Applying the conservative assumption that the 
modelled intensities should not exceed the limit set by observations of the 
XRB, halo parameter combinations can be constrained from Figure~\ref{fig2}.
Figure~\ref{fig2}a shows that for $T=T'$, $N_0\lesssim 5\times10^{-4}$ 
cm$^{-3}$ is required. Alternatively, $N_0=N'_0$ implies $T<0.25$ keV.
Figures~\ref{fig2}c--\ref{fig2}e show that for a realistic upper limit on 
$\beta$ of $\beta
\leq 1.0$, either $T \leq 0.3$ keV, $r_c \lesssim 10$ kpc, or $N_0 \lesssim
7\times10^{-4}$ cm$^{-3}$ is required. Figures~\ref{fig2}b and 
\ref{fig2}f show that
choosing $r_c\geq 20$ kpc implies either $T<0.25$ keV or $N_0\leq 6\times
10^{-4}$ cm$^{-3}$. We thus see that realistic choices of $\beta$ and $r_c$
strongly hint at either $N_0'$, $T'$, or both being too large. 
Although model intensities are thus incompatible with XRB observations, no
firm constraints can be put on halo parameters without further assumptions.

\begin{figure}
\epsscale{.6}
\plotone{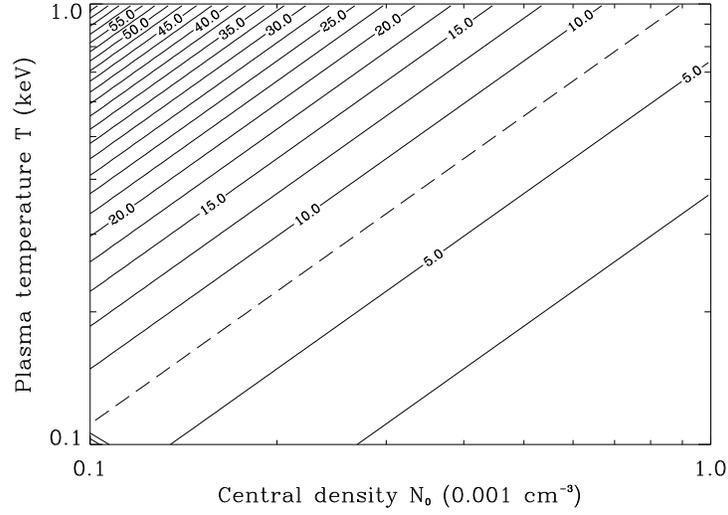}
\caption{Estimated central cooling time (upper limits, 
Gyr) of the LGH as function of central density $N_0$ and temperature $T$,
assuming the cooling function of Gehrels \& Williams (1993). Dashed line 
represents the condition $t_c=t_H/2$.\label{fig3}}
\end{figure}

\begin{figure}
\epsscale{.6}
\plotone{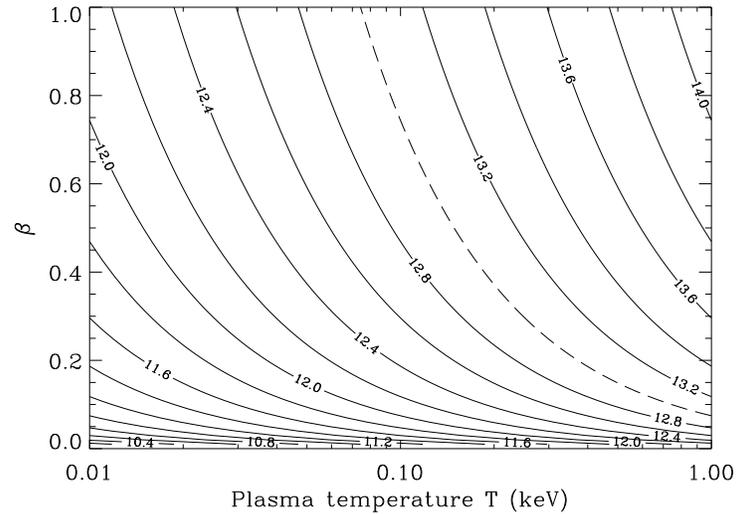}
\caption{Estimated total gravitating mass of the 
Local Group in units of log$(M_{grav}/M_{\odot}$) as function of global 
plasma temperature $T$ and $\beta$. Dashed line represents a conservative 
upper estimate on LG mass of $10^{13}$ M$_{\odot}$.\label{fig4}}
\end{figure}

More restrictive is the information that can be extracted from 
Figures~\ref{fig3} and \ref{fig4}. Our cooling stability 
requirement ($t_c>t_H/2$) restricts $T$ to above 0.1 keV for 
all $N_0 \geq 10^{-4}$ cm$^{-3}$ (Fig.~\ref{fig3}).
At $N_0=N_0'$ we require $T>1.1$ keV, while fixing $T$ at $T'$ implies 
$N_0<0.9\times 10^{-3}$ cm$^{-3}$. Both possibilities were ruled out above, 
but a stronger constraint on $N_0$, 
however, is provided by taking into account the estimated LG gravitating mass 
(Fig.~\ref{fig4}). Even assuming a conservative upper 
limit of $M_{grav}=10^{13} M_{\odot}$, Figure~\ref{fig4} shows that $T$
cannot exceed
0.5 keV for reasonable values of $\beta$, i.e.\ $\beta \gtrsim 0.15$ (0.6 keV
for $\beta>0.1$). As this
result is independent of $N_0$, we subsequently conclude from 
Figure~\ref{fig3} that $N_0<5\times 10^{-4}$ cm$^{-3}$. This conclusion 
is unaffected by any choice of $r_c \leq 300$ kpc, a core radius typical for 
clusters of galaxies. Given our assumptions, $N_0\lesssim 0.5N_0'$ and 
$T\lesssim 0.5T'$ is thus required for all $\beta >0.1$.
Notice moreover the very important fact that we are actually being 
conservative in two senses, since we take
the LG mass to be $<10^{13} M_{\odot}$ and {\em at the same time} allow
$\beta$ to be as small as $\sim 0.1$. Strictly requiring $\beta=0.5$ as 
representative of 
observed poor groups actually implies $T<0.15$ keV from 
Figure~\ref{fig4}, which by means of Figure~\ref{fig3} translates 
into $N_0 \lesssim 10^{-4}$ cm$^{-3}$. This result applies within 
$r_c\lesssim 400$ kpc. For comparison with previous models we also note that
$\beta=2/3$ implies $T<0.1$ keV with the same density constraint.

We note for completeness that although little evidence for radial plasma 
temperature declines in intragroup media is present 
\cite{mulc1996b,pild1996,pede1997}, the impact hereof in
our model would be to lower the above mass estimate and thereby loosen the 
constraints on $N_0$ and the central temperature $T_0$. 
Assuming, e.g., $T(r)=T_0r^{-\psi}$, we require 
$\psi>0.15$ at $T_0=T'$ for $M_{grav}$ to be $\leq 10^{13} M_{\odot}$, 
corresponding to a factor $\sim 3$ in temperature decrease from halo center 
to cut-off radius. We also note that
numerical simulations of clusters of galaxies show that
mass estimates by means of the general equation~(\ref{eq,mass}) are 
reasonably robust towards deviations from hydrostatic equilibrium and 
spherical symmetry \cite{schi1996}. Other cluster simulations 
(Evrard, Metzler, \& Navarro 1996) further show that if using an isothermal 
$\beta$-profile with equation~(\ref{eq,mass}), the resulting mass estimator 
--- applied here --- is nearly
unbiased, showing a standard deviation $\sigma \lesssim 50$\% for the region
within which the estimated overdensity of the halo is $\geq 100$. For a 
present--day halo of the initial model 
parameters, this range corresponds to $r\lesssim 1.0$ Mpc, assuming 
$H_0=65$ km s$^{-1}$. Thus, while the 
result of Evrard et al.\ (1996) may not directly support our conclusion, it 
at least
suggests that our very conservative choice of Local Group mass and of allowed 
range of $\beta$--values ensures that our limit on $T$ is both reasonable and
firm.

\section{A REVISED MODEL}\label{sec,newmodel}

Investigating model properties with the down-scaled values of 
$N_0=5\times 10^{-4}$ cm$^{-3}$ and $T=0.5$ keV as determined in 
\S~\ref{sec,discus}, this section considers the resulting model 
intensities, the prospects for detecting UV absorption lines from the halo 
gas, and the effect of the halo on the CMB.

\subsection{X-ray Intensities}\label{sec,newint}

To examine whether halo parameters can be constrained any further, we again
plot model intensities towards M31, now using the derived upper limit on
$N_0$ and $T$ as input values while maintaining the value of remaining 
parameters. We emphasize once more that at this
temperature, {\em ASCA} model intensities are only $\sim 60-70$\% those of 
{\em ROSAT},
the exact fraction depending on the choice of $Z$. 
We now also apply a more realistic but less conservative value for the
allowed model intensities, based on the fact that at least $70-80$\% of the 
0.5--2 keV cosmic XRB in the direction of the Lockman Hole has been resolved 
into discrete sources 
\cite{hasi1998} (the term {\em cosmic} here referring to the emission 
originating outside the Local Group). Hence, allowing for model intensities 
of 12 keV cm$^{-2}$ s$^{-1}$ sr$^{-1}$ keV$^{-1}$ is
perhaps a too conservative approach, since most of this 1 keV intensity
would surely originate outside the Local Group. We henceforth adopt a model 
normalization of 50\% of the initial value, i.e., 
6 keV cm$^{-2}$ s$^{-1}$ sr$^{-1}$ keV$^{-1}$. Even in the presence of small 
variations in the cosmic XRB over the sky, this should still be a conservative
estimate.

\begin{figure}
\epsscale{1.0}
\plottwo{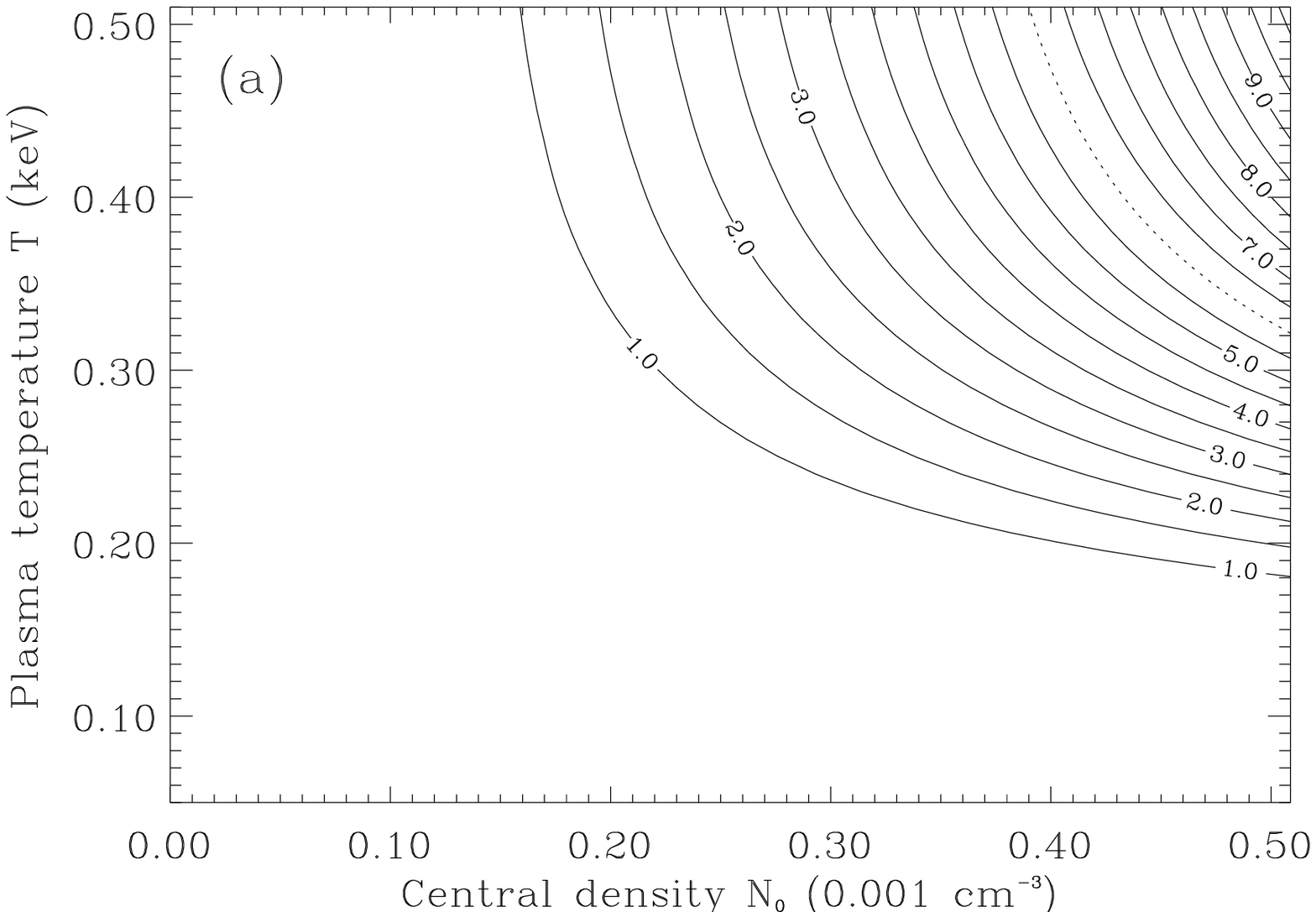}{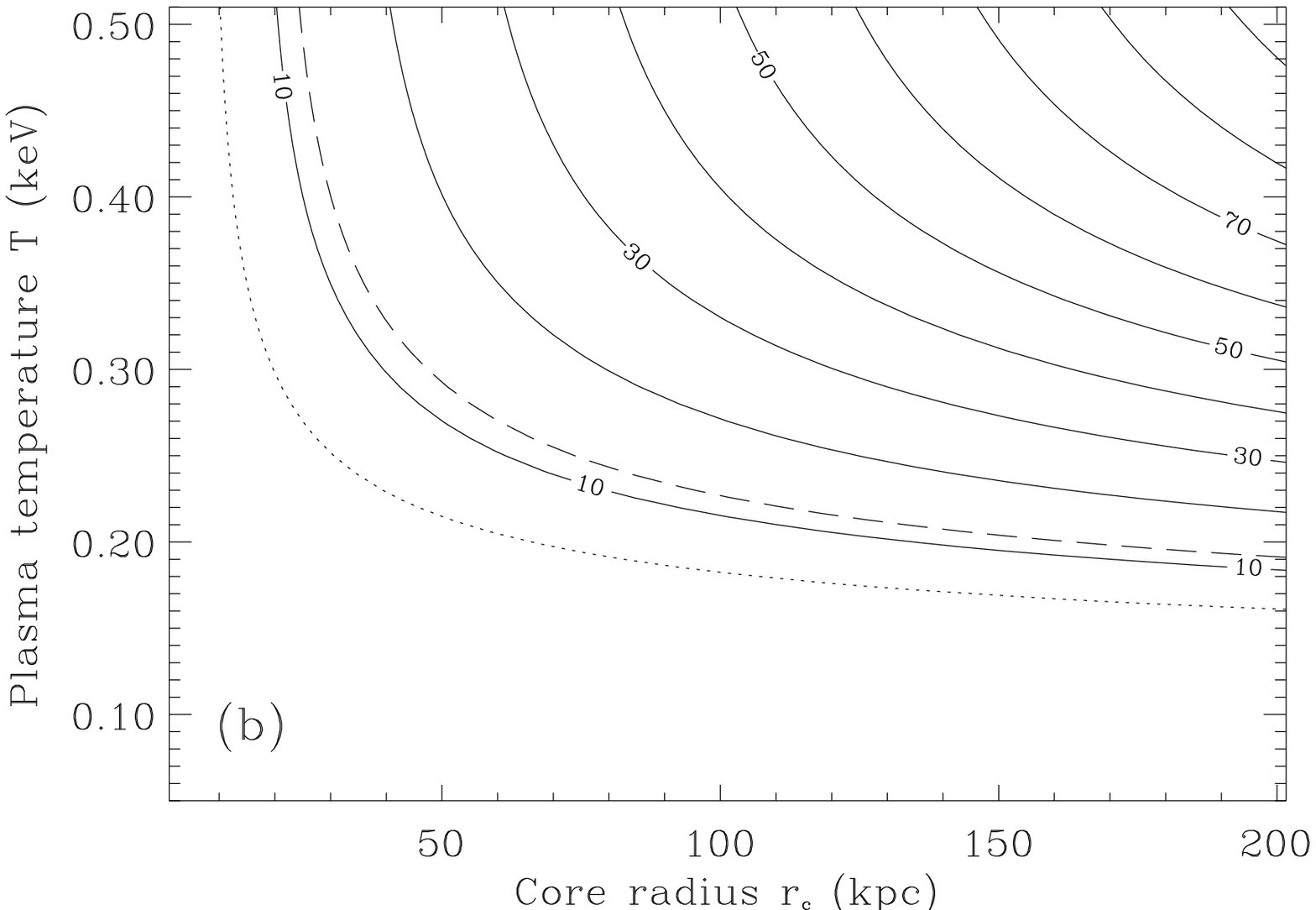}
\epsscale{2.22}
\plottwo{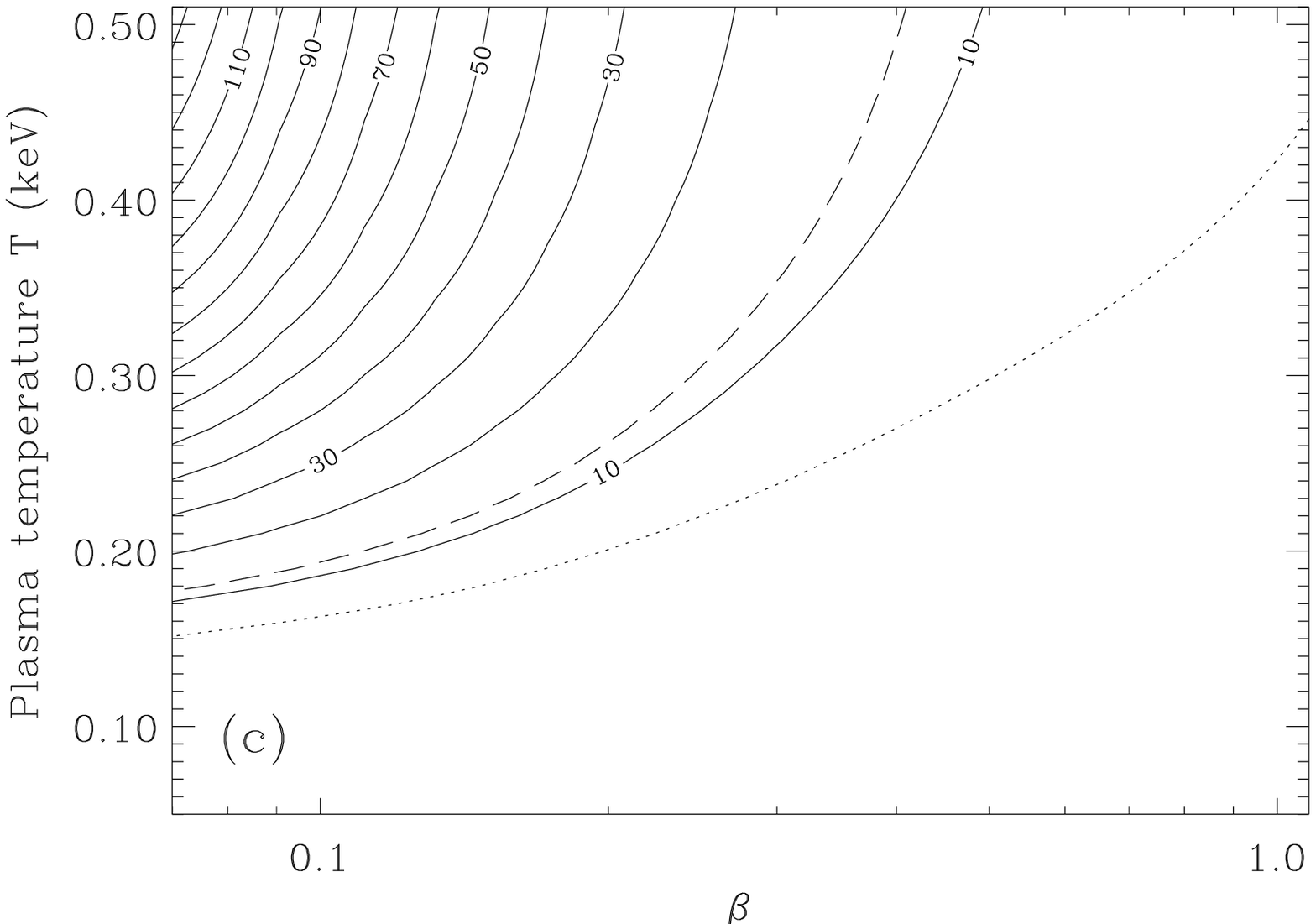}{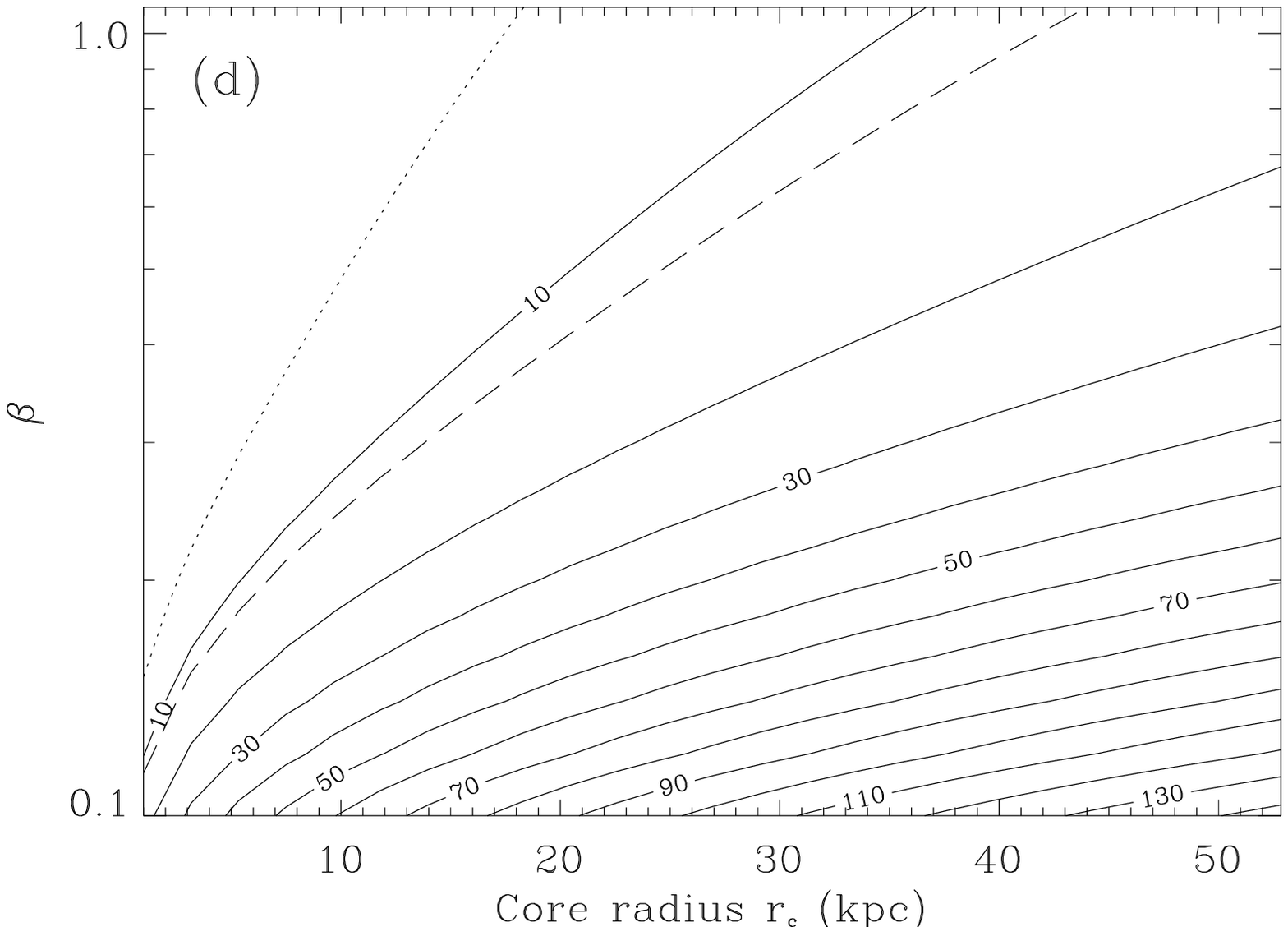}
\epsscale{4.94}
\plottwo{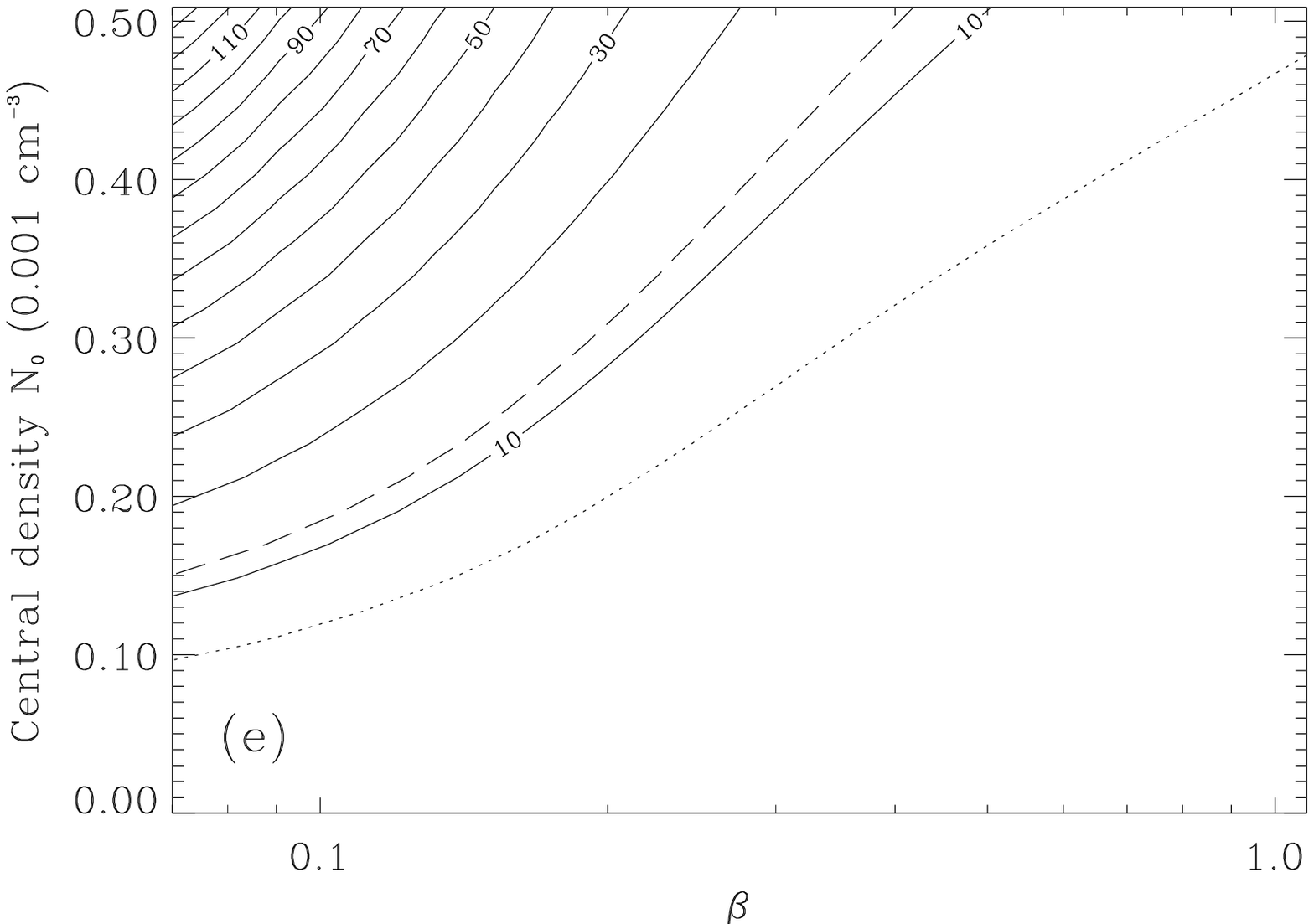}{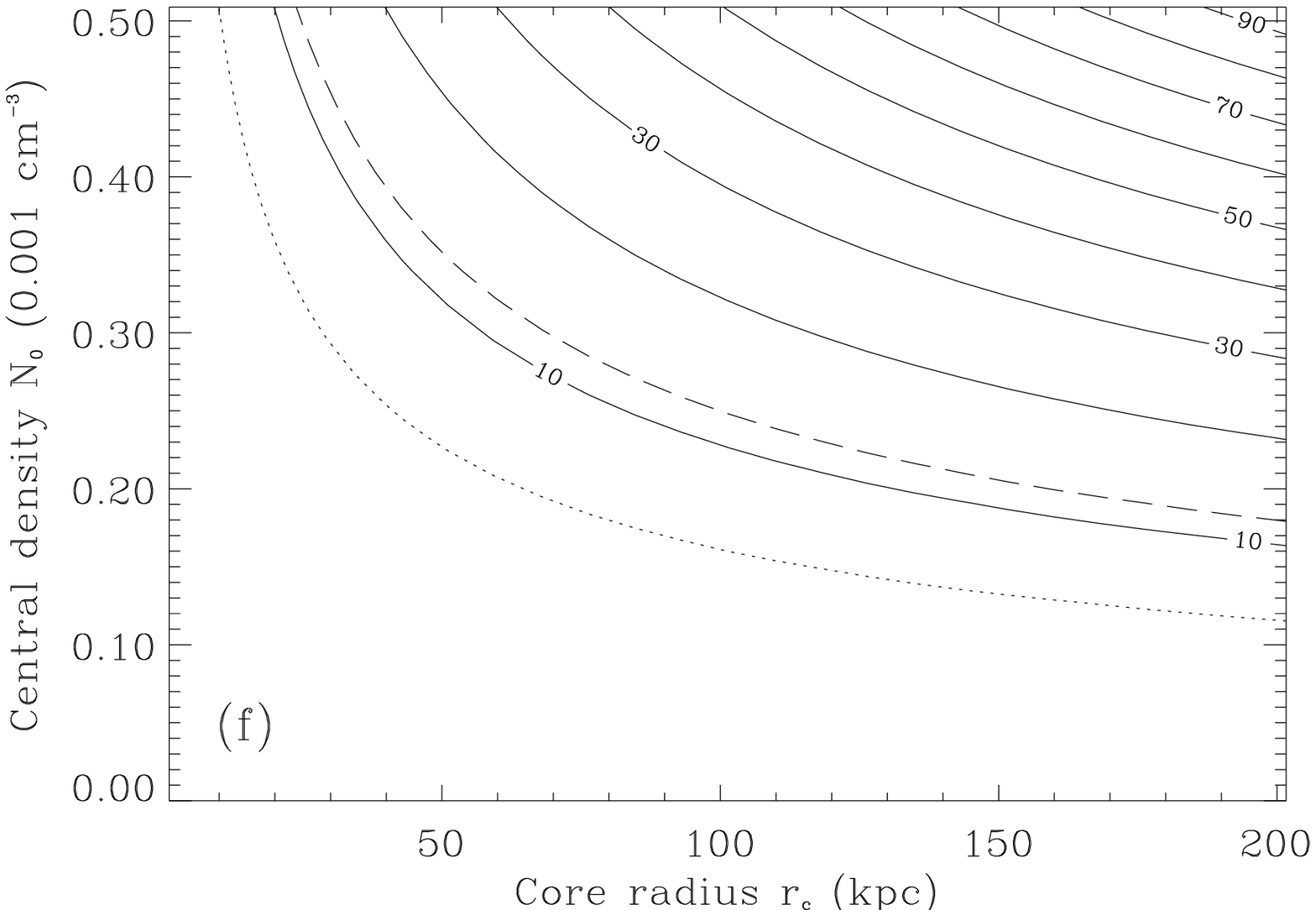}
\caption{As Figure~\ref{fig2} but for the revised model with 
$T=0.5$ keV and $N_0=5\times10^{-4}$ cm$^{-3}$. Dotted line marks a revised
estimate of the model normalization of 6 keV cm$^{-2}$ s$^{-1}$ sr$^{-1}$ 
keV$^{-1}$.\label{fig5}}
\end{figure}

Model results are shown in Figure~\ref{fig5}, including the initial 
(12 keV cm$^{-2}$ s$^{-1}$ sr$^{-1}$ keV$^{-1}$ --- dashed line)
and revised (6 keV cm$^{-2}$ s$^{-1}$ sr$^{-1}$ keV$^{-1}$ --- dotted line) 
XRB limits on model intensities.
\placefigure{fig5}
\notetoeditor{Please join Figures ``\ref{fig5}a''--''\ref{fig5}f'' 
into ONE two-column three-row figure, i.e.\ a
mock-up exactly equivalent to Figure~\ref{fig2}.}
Given the adopted values of $\beta$ and $r_c$, Figure~\ref{fig5}a shows 
that for our revised model intensity normalization either $N_0$ or $T$ has to
be scaled down at least a further $\sim20$\% to be compatible with 
observations. Figures~\ref{fig5}b and \ref{fig5}f show that $r_c\geq 
20$ kpc requires $T<0.3$ keV or $N_0<4 \times 10^{-4}$ cm$^{-3}$, while 
fixing $N_0$ and $T$ at their upper limits implies $r_c<10$ kpc. 
Figures~\ref{fig5}c--\ref{fig5}e show that for a low-end value of
$\beta=0.3$ (inferred for some groups), either $T<0.2$ keV, $N_0<2.5\times
10^{-4}$ cm$^{-3}$, or $r_c \lesssim 5$ kpc follows.
Finally, Figure~\ref{fig5}d shows that all realistic values of $\beta$
require $r_c\lesssim 20$ kpc. Thus, we cannot ``save'' the model by 
invoking a steeper halo density distribution without {\em at least} also 
decreasing the model core radius. As $r_c=20$ kpc is already a rather small
value compared to that derived for many other groups, this suggests that
our upper limits on $N_0$ or $T$ should be further decreased.
Conclusively, the model is now marginally consistent with the 1 keV 
normalization of the XRB but inconsistent with our revised intensity 
requirement, the reason probably being that both
temperature and central density still exceed the actual values. Further
narrowing in on the model halo parameters is, however, not feasible in terms 
of resulting X-ray intensities. From better estimates of the XRB
intensity (in particular the fraction of which is of non-cosmic origin) 
towards the LGBC,  
more precise ---and perhaps stronger--- constraints could be easily put 
forward.

\subsection{Prospects for UV Detection}

As is apparent from the previous discussion, the direct detection of X-ray
emitting gas associated with a Local Group halo is exceedingly difficult.
Since many galaxy groups dominated by spiral galaxies, such as the Local 
Group, could possess gas so cool and tenuous that it has yet escaped X-ray 
detection, it is natural to investigate whether such gas would
be detectable at longer wavelengths, e.g.\ below the {\em ROSAT} frequency 
limit of $\sim 0.1$ keV (i.e.\ $\lambda \gtrsim 100$ \AA). 
Specifically, at the relevant temperatures for Local Group gas 
($T<0.5$ keV), some line emission as well as absorption of externally 
originating radiation is 
expected to occur in the far- and extreme-UV bands, some of which, at least 
in principle, would be detectable with the {\em Far Ultraviolet Spectroscopic 
Explorer} ({\em FUSE}, covering the range $\sim 900-1200$ \AA), and the 
Space Telescope Imaging Spectrograph (STIS, covering a far-UV range of
$\sim 1150-1700$ \AA) aboard the {\em Hubble Space Telescope}. In addition 
to our model results so far, detecting such lines from LGH plasma could also 
provide potentially useful constraints on the LGH, the prospects of which 
will be briefly considered in the following. 

At the relevant plasma temperatures, and apart from lines of neutral hydrogen,
{\em FUSE} has the ability to observe the O {\sc vi} resonance doublet at 
rest-frame wavelength $\lambda \lambda = 1031.93, 1037.62$, while the 
{\em HST} STIS can observe the doublets of N {\sc v} and C {\sc iv} at 
$\lambda \lambda = 1238.82, 1242.80$ and
$\lambda \lambda = 1548.20, 1550.77$, respectively. Considering absorption, 
the O {\sc vi} doublet is the most interesting for our purposes, as it is 
expected to be the strongest metal line. In addition, in the wavelength range
covered by {\em FUSE} and {\em HST}/STIS, O {\sc vi} is the ion species 
least likely to be produced by photoionization, making it a prime indicator of
collisionally ionized plasma.
The possibility of detecting ---in the far-UV spectra of distant active 
galactic nuclei and quasars--- these and other higher-ionization absorption 
lines induced by gas in other spiral-dominated groups has been discussed by 
Mulchaey et al.\ (1996a). In addition, AGN/QSO far-UV absorption lines have 
also been used extensively to study the 
amount and spatial distribution of O {\sc vi} and other ions and atoms in the
Galactic 
halo from {\em FUSE} \cite{oege2000,sava2000a} and {\em HST} \cite{sava2000b} 
observations. Even extragalactic features such as the Magellanic Stream and
certain high-velocity clouds (HVC's) have been investigated in this manner 
\cite{semb2000}, indicating the potential of the method for LGH studies.

\begin{figure}
\epsscale{.6}  
\plotone{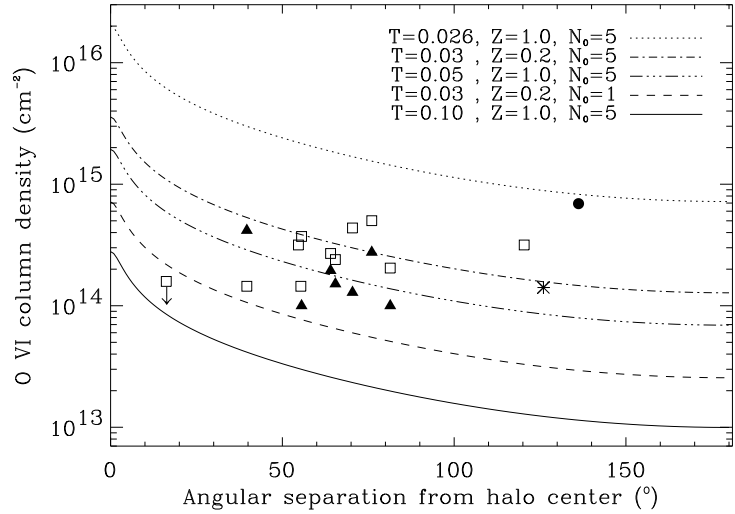}
\caption{O {\sc vi} column density of the Local Group halo for various 
assumed plasma temperatures (keV), metallicities ($Z/Z_{\odot}$), and central
electron densities ($10^{-4}$ cm$^{-3}$) as a function of
angular separation from the adopted halo center. Data points are
absorption line values; squares 
are from Savage et al.\ (2000) (downward arrow indicates an upper limit), and
triangles are from Sembach et al.\ (2000) along the same lines of sight --- 
see text for details. The
asterisk is an emission--line estimate towards M87 by Dixon et al.\ (2001),
while the filled circle is a result obtained by Hurwitz et al.\ (1998) using
the {\em ORFEUS II} satellite. Error bars have been omitted for clarity. 
\label{fig6}}
\end{figure}

Although largely attributed to the Galactic interstellar medium and halo,
some component of the observed O {\sc vi} absorption profiles and derived 
column densities could principally contain a contribution from LGH gas, 
despite the fact that this line is most prominent in relatively cool gas
($T\lesssim 0.1$ keV). 
To investigate this, we plot in Figure~\ref{fig6} the model halo 
value of O {\sc vi} column density $N($O {\sc vi}) for a few combinations of 
$T$, $Z$, and $N_0$ as a function of distance from the adopted LGH center.
Model predictions are calculated using a Raymond-Smith (1977; this version
last updated 1991 by J.C.\ Raymond) plasma code, and also plotted are 
observationally deduced column densities. These include 
the (presumably mainly Galactic) values derived by Savage et al.\ (2000a) by 
considering the low--velocity portion of absorption line profiles found
along 11 lines of sight, and the values of Sembach et al.\ (2000) who 
correspondingly considered the high--velocity part of these profiles 
(velocities $< -100$ km s$^{-1}$, if present) and were able to relate
this high-velocity gas to H {\sc i} HVC's in six of the seven high-velocity
cases (meaning that their derived column densities in these cases are extreme
upper limits to the presence of high-velocity gas in the Local Group in these
directions). 

\placefigure{fig6}

Firstly, we note that although the modelled column 
densities vary by a factor of $\sim 30$ across the sky, the data points in 
Figure~\ref{fig6} do not 
show any systematic variation with separation from the adopted LGH center, 
suggesting that they do indeed reflect mainly Galactic values. 
Further, the revised model halo 
($T=0.5$ keV, $Z=0.2Z_{\odot}$, $N_0=5\times 10^{-4}$ cm$^{-3}$) has not been 
included in the plot, as it
lies some 4 orders of magnitude below the low-value data points determined
from observations. Although this is not surprising in light of the previous
conclusion, we initially invoke the extremely conservative
requirement that $N($O {\sc vi})  of the LGH should not exceed observed 
values. Obviously, for all $T>0.1$ keV (given that $Z=Z_{\odot}$ 
is a likely extreme upper limit on halo metallicity), no further constraints 
can then be readily imposed on the model. 
Only for temperatures below this value it is 
possible to rule out certain combinations of $Z$ and $N_0$ for fixed $r_c$
and $\beta$. For instance, at the naively estimated virial temperature of the
halo of $\sim0.03$ keV (\S~\ref{sec,model}) it would be
possible to constrain $N_0$ by a factor of a few. At $T\simeq0.026$ keV, the 
O {\sc vi} abundance peaks, but a ``hot'' LGH at this temperature would need
a very low central density in order to fulfill our cooling stability 
requirement, in particular for large values of $Z$. The fact that even a 
single data point is close to this hypothetical upper 
O {\sc vi} limit in Figure~\ref{fig6} lends further support to the assertion 
that the O {\sc vi} absorption is primarily Galactic. 

We are thus immediately led to conclude that any attempt to use
the observed O {\sc vi} column densities as a further constraint on LGH 
parameters would require the halo temperature to be within the low and 
rather narrow range of $\sim0.03-0.1$ keV. Realistically, of course, 
the assumption that the observed $N$(O {\sc vi}) is entirely attributable to 
an LGH should be relaxed. However, given that a 
redshift of zero is determined for the absorbing gas, a large fraction 
of the observed absorption must at first be unambiguously assigned to the 
Galaxy in order to subsequently isolate and quantify the extragalactic 
contribution. The crucial problem in this context is the difficulty of 
obtaining an independent distance estimate for the gas in order to 
discriminate between the contributions from various absorbing layers along 
the line of sight (e.g., the Galactic interstellar medium and halo, the LGH,
HVC's). As noted by Mulchaey et al.\ (1996a), the thermal broadening of the 
lines (as reflected in the width of the Gaussian portion of the line 
profile) is probably insufficient as a discriminator, since gas in the 
Galactic halo could display temperatures similar to those discussed here for 
the LGH. Disentangling the contributions of multiple absorbers to the line 
width would also  ---at least--- require the lines to be fully spectrally 
resolved, which is probably at the limit of {\em FUSE} capability.
Yet another possibility in the context of O {\sc vi} would be to map 
the entire sky at far-UV wavelengths to determine the distribution of 
O {\sc vi} {\em emission} from gas in the Galactic halo.  
Since any O {\sc vi} emission from the LGH would be almost entirely
absorbed in the Galaxy, such a survey would primarily probe the O {\sc vi}
content of the Galactic disk and halo, allowing an estimate of the purely
Galactic column densities. Subtracting this from observed O {\sc vi} 
absorbing column densities, to which the LGH expectedly contributes, could 
provide an upper limit in any given direction on the contribution to the 
O {\sc vi} column density exterior to the Galaxy. At present, only a few 
firm results on O {\sc vi} emission are available from individual pointings, 
including those of Dixon et al.\ (2001) and Shelton et al.\ (2001).
However, a far-UV all-sky
survey using spectral imaging with arcminute spatial resolution is exactly
among the
goals of the proposed {\em SPEAR} mission ({\em Spectroscopy of Plasma 
Evolution from Astrophysical Radiation}; Edelstein, Korpela, \& Dixon 2000).

We note that the strong H {\sc i} lines expected  from a relatively 
cool collisionally ionized plasma in the {\em FUSE}--{\em HST}/STIS 
wavelength regions also provide a potentially powerful tool for
estimating LGH column densities. Savage et al.\ (2000b) present measurements
of the H {\sc i} Ly$\alpha$ absorbing column density in the Galactic disk 
and halo toward 14 QSOs at various Galactic latitude $b$. They find an 
average  value of 
$N$(H {\sc i}, Ly$\alpha$)$\mid \mbox{sin } b\mid =(1.29\pm 0.49)\times 10^{20}$ cm$^{-2}$, 
in reasonable agreement with H {\sc i} $\lambda21$ cm emission estimates.
This easily supersedes the values expected from our model LGH, which even
at temperatures as low as 0.01 keV only displays a H {\sc i} column density
of $\sim 10^{15}$ cm$^{-2}$ towards the LGBC, making it virtually impossible 
to place further constraints on, e.g., $N_0$ using this approach.

At higher and perhaps more relevant temperatures, other species producing 
strong line doublets, such as Ne {\sc viii}, Mg {\sc x}, or Si {\sc xii}, 
could be considered instead. Our revised model predicts central column 
densities for these ions of $1\times 10^{12}$, $2\times 10^{13}$, and 
$1\times 10^{14}$ cm$^{-2}$, respectively (increasing nearly two orders of 
magnitude for Ne {\sc viii} and Mg {\sc x} and a factor of $\sim 4$ for 
 Si {\sc xii}, if the temperature is lowered to 0.1 keV at the same $Z$), 
but their lines all fall in the rest--frame wavelength range of $500-800$ 
\AA$\:$ (see, e.g., Verner, Tytler, \& Barthel 1994), making them 
unobservable with {\em FUSE} or {\em HST}/STIS.  The {\em Extreme Ultraviolet 
Explorer} ({\em EUVE}) covered this wavelength region (with a spectral 
resolution $\gtrsim 1.3$ \AA$\:$  for $\lambda\gtrsim 500$ \AA) as did the 
BEFS spectrometer aboard the {\em ORFEUS/SPAS} space shuttle missions 
(resolution $0.1-0.2$ \AA$\:$ between 400 and 900 \AA; Hurwitz \& Bowyer 
1996). To our knowledge, however, estimates from these missions on the 
absorption line widths or column 
densities of the three ions species of interest are not available and would
for our purpose in any case suffer from the same potential systematics as in 
the O {\sc vi} case, although perhaps to a lesser extent. 
In summary, it does not appear to be feasible to impose further constraints
on a Local Group X-ray halo by applying available UV absorption line data. A 
more detailed account of the amount, distribution, and chemical composition of
Galactic gas in various directions is required for this approach to be 
fruitful.

\subsection{Distortion of the CMB?}

The presence of a Local Group halo is expected to affect the CMB through the 
Sunyaev-Zel'dovich effect.
The topic has previously been discussed by Suto et al.\ (1996),
Banday \& Gorsky (1996), Pildis \& McGaugh (1996), and Osone et al.\ (2000),
specifically in terms of the induced quadrupole anisotropy.

We may quantify the effect either in terms of the temperature
decrements $\Delta T_r$ seen in given directions relative to the CMB 
temperature $T_r$, or by the all-sky multipole 
anisotropies imprinted on the CMB by the halo. Beginning with the former 
approach we have
\begin{equation}\label{eq,S-Z}
\frac{\Delta T_r}{T_r}=y(xcoth(x/2)-4),
\end{equation}
where
\begin{equation}\label{eq,y}
y=\int_0^{\xi}{\frac{kT_e}{m_ec^2} \sigma_T N_e d\xi}
\end{equation}
is the Compton $y$-parameter. Here $T_e$ is the electron temperature, $\xi$ 
is the length of line of sight through halo gas,
$\sigma_T=(8\pi /3)r_e^2$ is the
Thomson electron scattering cross section (with $r_e = $ the classical
electron radius), and $x\equiv h\nu /(kT_r)$ (with $\nu = $ the frequency of
radiation). From this, the largest temperature decrement as seen towards
the adopted LGBC is only $|\delta T| \sim 0.5$ $\mu$K at $\nu=53$ GHz. This 
effect is too
small to be observable, let alone clearly identifiable, in any of the CMB maps
of Bennett et al.\ (1996) based on {\em COBE} DMR data.

It was suggested by Suto et al.\ (1996) that the LGH could induce a
quadrupole anisotropy $T_{2,SZ}$ in the CMB comparable to that seen be 
{\em COBE}, 4 $\mu$K $\leq T_{2,rms} \leq 28$ $\mu$K 
(95\% confidence; Kogut et al.\ 1996).
Following the approach of Suto et al.\ we expand
the CMB temperature variations in spherical harmonics to derive the 
LGH--induced monopole 
($T_{0,SZ}$), dipole ($T_{1,SZ}$), and quadrupole ($T_{2,SZ}$) 
anisotropies, imprinted on the CMB by a density distribution of the form of 
equation~(\ref{eq,beta}). For our spherically symmetric halo, this expansion 
becomes
\begin{equation}\label{eq,my}
\left( \frac{\delta T}{T} \right)(\mu) = \sum_{l=0}^{\infty}
a_{\ell} \sqrt{ \frac{2 \ell +1}{4\pi}} P_{\ell}(\mu) .
\end{equation}
Here $\mu=cos\theta$ measures the angular distance $\theta$ from the halo 
center, while $P_{\ell}$ are ordinary Legendre polynomials.
The $a_{\ell}$ are computed in the Rayleigh-Jeans regime using Suto's 
equation~(5),
\begin{equation}
a_{\ell}=-2\sqrt{\pi(2\ell+1)}\frac{kT_e\sigma_T}{m_ec^2}
\int_{-1}^{1}{\Sigma_0 P_{\ell}(\mu)d\mu} ,
\end{equation}
by means of the electron column density 
$\Sigma_0 = \int_0^{\infty}{N_e(r)dr}$, which for a $\beta$-profile is
\begin{equation}\label{eq,III}
\Sigma_0(\mu)=N_0 \int_0^{\infty}{\left(1+\frac{\xi^2-2x_0\mu \xi + x_0^2}
{r_c^2}\right)^{-3\beta/2}d\xi},
\end{equation}
with all parameters defined above. The latter integral can only be 
evaluated numerically, except in the case $\beta=2n/3$, ($n=0,1,2,$ \ldots). 
The multipoles are subsequently found from the identity
\begin{equation}
\frac{1}{4\pi}\sum_{\ell=0}^{\infty}(a_{\ell})^2\equiv 
\sum_{\ell=0}^{\infty} (T_{\ell,SZ})^2 , 
\end{equation}
which is Suto's equation (6). Results for the anisotropies induced by the 
initial and revised 
models are presented along with the corresponding {\em COBE} values in
Table~\ref{tbl-1}.

\placetable{tbl-1}

It is obvious from Table~\ref{tbl-1} that model-induced anisotropies are
far smaller than those measured by {\em COBE}. In particular, varying $r_c$  
anywhere in the interval 20--450 kpc at most increases $T_{2,SZ}$ by a factor 
of 10 which is still three orders of magnitude below the {\em COBE} value. 
It thus 
appears exceedingly unlikely that a LGH should contribute measurably to the
CMB quadrupole seen by {\em COBE}. This conclusion is in quantitative 
agreement with that of Osone et al.\ (2000) and consistent with the results
of Pildis \& McGaugh (1996) and Banday \& Gorski (1996).

We note that the $y$-parameter of both models in Table~\ref{tbl-1} are two 
orders of magnitude below the upper limit of $y<1.5\times 10^{-5}$ inferred 
from {\em COBE} data \cite{fixs1996}.

\section{CONCLUSION} \label{sec,conclude}

We can briefly summarize the conclusions as follows: \\

\noindent{1)} The modelled Local Group halo emission is very localized in 
the sky, suggesting that 
a LGH centered in the direction of M31 would be hard to distinguish 
observationally from the combined emission of M31 and the extragalactic 
X-ray background.\\
2) Applying typical values of morphological and physical parameters of
hot gas observed in other groups, the modelled intensities exceed those 
allowed by observations of the X-ray background. \\
3) Estimates of Local Group mass strongly suggest a temperature 
$T<0.5$ keV for an
isothermal halo in hydrostatic equilibrium. Within this temperature range, 
the stability of the halo against significant
cooling requires a central density $N_0<5\times10^{-4}$ cm$^{-3}$. These 
values apply regardless of the values of $\beta$-profile parameters $\beta$
and $r_c$, but resulting X-ray intensities then require $r_c\lesssim 40$ kpc.
A typical poor X-ray group value of $\beta=0.5$ 
requires $T<0.15$ keV and $N_0 \lesssim 10^{-4}$ cm$^{-3}$. 
For $r_c<60$ kpc, this density constraint is more restrictive than the 
(possibly underestimated) $T=0.3-1.2$ keV constraint on $N_0^2r_c$ of Osone 
et al.\ (2000), 
even though a shallower halo density profile is assumed here. \\
4) Using the derived conservative upper limits on $N_0$ and $T$
of $5\times 10^{-4}$ cm$^{-3}$ and 0.5 keV, respectively,
resulting X-ray intensities are marginally consistent with the 1 keV 
normalization of the X-ray background. Taking into account the fact that
the majority of the 1 keV background is comprised of individual sources at
cosmic distances, model results suggest that the upper limit on $N_0$ or 
$T$ should be decreased at least by $\sim 20$\%. An observational estimate 
of the non-cosmic background seen towards the Local Group barycenter could 
sharpen the constraints on halo parameters even further.  \\
5) The possibility of deriving further constraints on halo parameters from
the ability of the halo to produce absorption lines in the far-UV spectra of 
quasars is severely restricted to the case of halo temperatures $T<0.1$ keV, 
given currently available data on UV absorption lines and the absence of a
general method for estimating the distance to the absorbing gas. \\
6) The effect of the model halo on the Cosmic Microwave Background is entirely
negligible, both in terms of induced absolute CMB temperature variations and
in terms of multipole anisotropies. 

Of course, a true Local Group halo may be neither spherical, isothermal, or 
in hydrostatic equilibrium. But (1): We use a conservative upper Local Group
mass estimate, more than twice the largest value derived from recent 
observations; 
(2): The Local Group galaxy
distribution is indicative of significant mass subclustering, with 
little luminous mass associated with the central $\sim300$ kpc. That is, 
$N_0$ may be even lower than our estimates suggest, and we therefore 
believe that our limit on $N_0$ is rather firm. Further, the inference of a 
relatively 
low halo temperature is in accordance with estimates of the Local Group 
virial temperature as derived from the galaxy velocity dispersion. 
Conclusively, it therefore appears unlikely that the Local Group should 
possess any significant X-ray halo.
We note in passing that the integrated 0.1-2.0 keV luminosity of the model
halo is $2.8\times 10^{40}$ erg s$^{-1}$ ($1.8\times 10^{40}$ erg s$^{-1}$
for 0.5-2.0 keV), which is lower than typical X-ray 
luminosities inferred for X-ray groups of galaxies and also lower than typical
upper limits inferred for groups {\em not} detected in X-rays with 
{\em ROSAT} or {\em ASCA} \cite{mulc1996b}. 
Thus, we would not expect groups like the 
Local Group to have been detected by these satellites. This is consistent 
with the current observational status on small groups of galaxies 
\cite{zabl1998,zabl1999}.

\acknowledgments

We thank the referee for prompt and constructive comments.
Both authors acknowledge support by the Danish Natural Science Research
Council (SNF).

\clearpage

\clearpage

\clearpage

\clearpage

\clearpage

\begin{deluxetable}{cccc}
\footnotesize
\tablecaption{Modelled and Observed CMB Anisotropies. \label{tbl-1}}
\tablewidth{0pt}
\tablehead{
\colhead{Multipole} & \colhead{Initial Model} & \colhead{Revised Model} & 
\colhead{{\em COBE}}
}
\startdata
$T_{0,SZ}$ & 9.8 nK & 1.5 nK & $T_0=2.728\pm 0.004$ K\tablenotemark{a} \\
$T_{1,SZ}$ & 4.2 nK & 0.6 nK & $T_1=3.372\pm 0.014$ mK\tablenotemark{a} \\
$T_{2,SZ}$ & 2.5 nK & 0.4 nK & 4 $\mu$K $\leq T_{2,rms} \leq 28$
$\mu$K\tablenotemark{b} \\
\enddata

\tablenotetext{a}{Data from Fixsen et al.\ 1996}
\tablenotetext{b}{95\% confidence interval (Kogut et al.\ 1996)}
\end{deluxetable}

\end{document}